\documentclass[journal]{IEEEtran}

\usepackage{cite}
\usepackage{amsmath}
\usepackage{amssymb}
\usepackage{graphicx}
\usepackage{booktabs}
\usepackage{array}
\usepackage{microtype}
\usepackage{url}
\usepackage[hidelinks]{hyperref}
\hypersetup{pdftitle={A Threshold Model of Fault Propagation in AI-Native Software Ecosystems},pdfauthor={Daniel Russo}}

\begin{document}

\title{A Threshold Model of Fault Propagation in\\ AI-Native Software Ecosystems}

\author{Daniel~Russo%
\thanks{D. Russo is with the Department of Computer Science, Aalborg University,
2450 Copenhagen, Denmark (e-mail: daniel.russo@cs.aau.dk); ORCID
0000-0001-7253-101X.}%
\thanks{This work was supported by Innovation Fund Denmark under the Grand
Solutions programme, grant no.\ 4354-00006B, through the project Human-Centered
Adoption of Artificial Intelligence for Software Engineering in Denmark
(AI4SE1DK).}}

\markboth{Preprint, November 2026}%
{Russo: A Threshold Model of Fault Propagation in AI-Native Software Ecosystems}
\newcommand{\thisdocument}{preprint}

\IEEEtitleabstractindextext{%
\begin{abstract}
When many coding agents work in one repository at once, software engineers are left to judge
a collective through its parts. Review examines one change at a time, yet a change can
incorporate a fault introduced by another, and repairing the originating change then no longer
repairs the repository. To keep human judgment at the centre of agentic software development, we need
models of what agents do together. This paper joins complexity theory with software
evolution to build one. We model how a fault spreads among open changes as a branching
process over \emph{contention}, the relation between two changes that are open at the same
time and modify a common file. As with the reproduction number of an epidemic, spread sustains
itself once each faulty change passes the fault, on average, to more than one other. Each
repository has a reference threshold, the crossing probability per contention link
above which spread would sustain itself in a random graph with the repository's contention
degrees. We reconstructed contention among 33,105 agent-authored pull requests in 2,731
repositories. Thirteen repositories with reference thresholds below 5\% contained 42.5\% of
the changes. There, a fault that crossed one contention link in twenty would sustain spread
under the model. The structural conditions for propagation are in place where
much agent-authored work happens. Reviewing each change separately cannot reveal these conditions.
Engineers also need to know where concurrent work exposes a repository to spreading faults,
which they can compute from records they already keep.
\end{abstract}

}

\bstctlcite{IEEEexample:BSTcontrol}
\maketitle
\IEEEdisplaynontitleabstractindextext
\IEEEpeerreviewmaketitle

\section{Introduction}\label{sec:introduction}

\IEEEPARstart{M}{ore} is different. With this phrase, Anderson argued that ``at each level of complexity entirely new properties appear''~\cite{Anderson1972}. Collectives of AI agents now offer a case that is close to software engineering. In July 2026, roughly 1,200 agents that ran in separate sandboxes and were meant to be fully isolated from one another turned a shared package cache into a message board. They exchanged more than 70,000 messages and files, and about 700 of them went on to take part in an attack on an external service, which lay outside the scope of their tasks~\cite{METR2026}. None of this behaviour had been designed. The incident was extreme, but it raises a question that ordinary software development now shares. Coding agents submit changes to software repositories at scale. One public dataset records more than 930,000 pull requests authored by such agents across 116,211 repositories~\cite{AIDev2026}. \textbf{What do these agents do together that none of them does alone?}

Each change arrives with its own tests and is reviewed on its own terms. However, what a change does to the repository also depends on the other changes open at the same time, the files they share, and the order in which they are integrated. Software engineering has studied this dependence in human teams for decades~\cite{Gall1998, CataldoHerbsleb2013, Brun2011}. With coding agents, it may operate at a different scale, because one engineer can keep many agent changes open at once, and an agent can respond to a failing test without any engineer seeing the response. Review practice, in turn, is organised around the single change~\cite{RigbyBird2013, Sadowski2018}. It rests on an assumption that is rarely stated: \emph{a repository is correct when each of its changes is correct}. Work on multi-agent risk has already questioned the general form of this assumption, noting that ``a collection of safe agents does not guarantee a safe collection of agents''~\cite{GradientInstitute2025}. This paper challenges it for AI-native software ecosystems. We start from the repository, because this is where the collective behaviour of agents leaves records that can be measured.

This problem also concerns the role of the engineer. Engineers remain accountable for the software that agents produce, and review is where they exercise their judgment. If failures arise from what agents do together, this judgment is applied to the wrong unit, however careful each review is. We see the task of restoring it as central to agentic software engineering~\cite{Hassan2025} in the coming decade. \textbf{The task calls for a new way of thinking, which joins software engineering with complexity science}, the study of how collective behaviour arises from interacting parts~\cite{Anderson1972, Holland1995}. It also calls for models that tell engineers where a collective of agents needs their attention. Such a model has to start from a measurable mechanism, and this paper proposes one.

We study when a fault that enters one change stays there, and when the changes open beside it carry the fault further. Several changes can fail together in two ways. In the first, they all depend on one faulty origin. These failures are shared and disappear once the origin is repaired. In the second, another change \emph{incorporates} the faulty behaviour into its own content. The fault then survives the repair and can be passed on (Section~\ref{sec:object:episode} gives a hypothetical example). We propose that how far a fault travels depends on the set of open changes as much as on any agent within it. Two repositories whose agents are equally careful could differ in the reach of a single fault, because the routes between their changes differ. If those routes can be recovered from development records, a repository's history can show where a fault could travel far before any spread occurs. The mechanism applies to human-authored changes as well. Nevertheless, as agents may make it pressing, we measure contention among agent-authored changes.

We begin from the simplest relation that development records support. Two changes are in \emph{contention} when they are open at the same time and modify at least one common file. Contention is broader than a merge conflict, since a change can build on another's state through edits that merge cleanly. It can be reconstructed from any repository's history and provides a fault with a route. A change that carries a fault can pass it to a change that shares a file with it, which can pass it on in turn. However, contention records only the opportunity. Whether a fault actually crosses a contention link is a separate event, which the model represents by one parameter, the crossing probability $q$. Spread then follows the logic of the reproduction number of an epidemic. It sustains itself once each faulty change passes the fault, on average, to more than one other. Each repository therefore has a \emph{reference threshold}, which is the value of $q$ above which spread would sustain itself in a random graph with the same contention degrees as the repository. The threshold depends on how unevenly contention is distributed as well as on its average, because a fault that follows a link tends to arrive at a heavily contended change. In practical terms, it states how rarely faults may cross between contending changes before concurrent work starts to amplify them.

We reconstructed historical contention among 33,105 agent-authored pull requests in 2,731 public repositories. We found that reference thresholds differ across repositories by more than two orders of magnitude, and that agent-authored work is concentrated where they are lowest. \textbf{Thirteen repositories with reference thresholds below 5\% contained 42.5\% of the changes}, and a single vendor's coding agent authored 95.9\% of that subset. Under the model, a fault that crossed one contention link in twenty would be enough to sustain spread there. The structural conditions for propagation are therefore already present in the repositories where much agent-authored work takes place.

We make three contributions in this paper. \emph{Conceptually}, we define the units of a collective failure in software development. These are contention, the relation that provides a fault with a route, and incorporation, the step that allows a fault to survive the repair of its origin. \emph{Theoretically}, we connect percolation theory to software evolution through a model that assigns each repository a reference threshold, computed from its history. The model yields three testable propositions and the observations that separate fault propagation from review overload. \emph{Empirically}, we show that reference thresholds vary by more than two orders of magnitude across 2,731 repositories, and that agent-authored work is concentrated where thresholds are lowest. These findings motivate a research agenda on the collective behaviour of coding agents.

The remainder of this paper is organised as follows. Section~\ref{sec:related} reviews related work. Section~\ref{sec:object} defines AI-native software ecosystems and the fault-propagation mechanism. Section~\ref{sec:threshold} develops the threshold model, and Section~\ref{sec:corpus} applies it to a corpus of agent-authored changes. Section~\ref{sec:predictions} derives the propositions. Section~\ref{sec:implications} discusses the implications and the threats to validity, Section~\ref{sec:future} presents the research agenda, and Section~\ref{sec:conclusion} concludes.

\section{Related Work}\label{sec:related}

Three lines of software engineering research are relevant to our model. Most of this research concerns pairs of changes made by human developers. The first line shows that changes can depend on one another even when the code contains no explicit dependency between them. Files that change together in release histories are logically coupled, although the architecture does not document this coupling~\cite{Gall1998}. Change-propagation prediction learns which entities change together, so that an engineer who modifies one can be warned about the others~\cite{HassanHolt2004}. Unmet coordination needs, including concurrent work on the same files, are associated with integration failures~\cite{CataldoHerbsleb2013}, and changes can merge cleanly and still conflict in behaviour~\cite{Mens2002}. The second line builds tools that detect such conflicts early. Palant\'{\i}r makes parallel changes in other developers' workspaces visible as conflicts form~\cite{Sarma2012}. Crystal merges open branches speculatively and tests the result, which detects conflicts that a textual merge misses~\cite{Brun2011}. The third line avoids overlap or tests it before integration. Cassandra schedules tasks so that developers who are likely to touch the same files do not work at the same time~\cite{KasiSarma2013}. At industrial scale, a submit queue tests pending changes against one another before any of them lands~\cite{Ananthanarayanan2019}. Review practice, however, is organised around the single change. Changes are small, frequent, and quickly reviewed~\cite{RigbyBird2013, Sadowski2018}, and understanding the change is the difficulty that reviewers report most often~\cite{Bacchelli2013}.

Taken together, this work shows how changes that are correct on their own can combine into failures. However, it acts on one pair or one queue of changes at a time, or it prevents overlap before it occurs. Our threshold model builds on this work by asking a question about the whole population of changes. When does the set of concurrently open changes transmit faults strongly enough for one fault to keep moving? Early measurements of agent-authored work make this question pressing. They report conflicts between concurrent agent changes~\cite{Xu2026}, quality and architectural patterns particular to generated code~\cite{Zhu2026}, and repository-level variation in how agent contributions are handled~\cite{RussoGovern2026}.

Software has also been studied as a system with behaviour of its own. Mogul~\cite{Mogul2006} traced failures to interactions among correct components, and Perrow~\cite{Perrow1984} argued that tightly coupled systems fail through interacting failures. Lehman's laws state regularities of the evolving system as a whole~\cite{Lehman1996, Herraiz2013}. Research on self-adaptive systems built the control machinery for systems that manage themselves at runtime~\cite{KephartChess2003, deLemos2013, Weyns2020}, and it now places generative models inside the adaptation loop~\cite{LiWeyns2024}. This machinery assumes a signal that tells the system when to adapt. Our paper concerns where such a signal could come from in a development ecosystem.

So far, most evidence on multi-agent AI failure concerns agents that exchange messages to complete a task within minutes. Cemri et al.~\cite{Cemri2025} organise such failures in the MAST taxonomy, and risk analyses of governed deployments describe the same setting~\cite{GradientInstitute2025}. Xie et al.~\cite{Xie2026} model error cascades along the message channel. They derive an amplification criterion from the spectral radius of the message graph's adjacency matrix, and they report that a governance layer on that channel prevents an error from hardening into false consensus in at least 89\% of runs. In those runtime systems, contention means simultaneous access to shared memory or tools within one task~\cite{Khan2026}. In our study, contention is a slower relation between whole changes that stay open for days over a common file. We therefore take from this evidence the form of the question, namely whether a propagation criterion computed on an interaction graph predicts collective failure. The graph, the time scale, and the rates are different.

The mathematics that we use comes from statistical physics and network science, where it is well established. These fields study how local interactions among many connected elements produce the behaviour of a whole system, and they address the question of when a local event remains local and when it spreads. This is the question that we ask about faults. Percolation theory, in particular, locates the point at which local transmission on a graph becomes global~\cite{Newman2002, KarrerNewmanZdeborova2014}. Within software engineering, vulnerability propagation through package dependency networks has been measured empirically~\cite{Decan2018}. Pedreschi et al.~\cite{Pedreschi2025} study how humans and AI influence one another through a training loop that closes with each model release. In contrast, the influence that we study runs within hours, through code that agents share. Our contribution is to connect these bodies of work. We apply percolation theory to the concurrently open change and the files it shares. This lets us compute, from a repository's own records, the crossing probability per contention link above which a random graph with its contention degrees would sustain fault spread. To our knowledge, this connection has not been made before.

\section{AI-Native Software Ecosystems}\label{sec:object}

\subsection{System Boundary and Units of Analysis}\label{sec:object:definition}

Manikas and Hansen define a software ecosystem as the interaction of actors on a common technological platform that results in software~\cite{Manikas2013}. Earlier ecosystem and systems-of-systems models assume a person behind every change~\cite{Jansen2009, Maier1998}. We adapt the definition to the case in which an agent can submit a change without one. An \emph{AI-native software ecosystem} consists of the coding agents, software engineers, and automated tools that work on one repository and its continuous integration and deployment pipeline, together with the governance that sets review policy, approval workflow, and the division of responsibility between agents and people. The repository and its pipeline are the operational boundary throughout. Package ecosystems such as npm lie outside it. They connect the software that repositories produce, whereas the ecosystem studied here is the set of actors producing software in one repository. A runtime multi-agent system completes a task within minutes and fails through its messages. In contrast, a development ecosystem runs for weeks and fails at integration, when changes conflict~\cite{Xu2026}, introduce unintended coupling~\cite{Zhu2026}, exceed review capacity~\cite{Sadowski2018}, or pass a regression from one change to another. Our model concerns the last of these failures.

The nodes of our model are changes. The corpus requires us to distinguish a change from the agent behind it. A \emph{change} is a pull request with its files and its open interval. An \emph{artefact} is a file a change modifies, and a \emph{fault site} is a location in an artefact that can be correct or faulty with respect to its specification. An \emph{account} is the identity under which a change was opened, a \emph{product} is the coding-agent system credited with authoring it, and an \emph{agent instance} is one running copy of a product working on one task. The corpus of Section~\ref{sec:corpus} records accounts and products but not instances, and three of its five products author changes through the invoking engineer's account~\cite{AIDev2026, RussoGovern2026}. The theory concerns a repository receiving changes from several agent instances at once. However, the corpus observes contention among agent-authored changes without recording how many instances produced them.

\subsection{Interaction Channels}\label{sec:object:connections}

Agents in such an ecosystem can be connected through shared code, messages, shared resources such as caches and build artefacts, and delegation or approval, when one agent's output becomes another's task or a review decision admits a change. For each connection, we treat three attributes separately. These are whether an agent was permitted to use it, whether it actually used it, and whether it created or repurposed it. A connection that is missing from the records is unknown, and it should not be treated as absent. The OpenAI and Hugging Face incident of July 2026 (Section~\ref{sec:introduction}) is an example, since a shared package cache became an unsanctioned message board~\cite{METR2026}. \textbf{No version history would show that a cache had become a message channel.}

We measure shared code because version history records which files each change modified. The corpus used here limits that record in two ways. First, it contains agent-authored changes only. Second, it keeps each change's final file list, the union of files modified in its recorded commits, without the order in which they were touched (Section~\ref{sec:threshold:temporal}).

We compared the two channels that the corpus records. Coding-agent accounts posted 4,674 messages on changes authored by others, against 42,600 contending pairs, a ratio of 0.110 (95\% cluster-bootstrap interval across repositories, 0.0645--0.233). Contending pairs thus outnumbered agent-to-agent messages by about nine to one. Our test specification took a ratio below 0.1 as support for shared code as the dominant channel. The estimate lies just above that bound, so we read the result as suggestive. The message count is a lower bound, since messages that an agent posts through an engineer's account are attributed to the engineer (Supplement~S5).

\subsection{Fault-Propagation Mechanism}\label{sec:object:episode}

We constructed the following hypothetical episode to illustrate the mechanism and the scope of the model. A billing module, \texttt{period.py}, decides which days belong to a billing period, and its specification says that a period includes both its first and its last day. Change $e_1$ modifies \texttt{period.py} to rewrite the helper \texttt{in\_period} and, in doing so, makes the last day exclusive. Its tests check dates in the middle of a month, so the fault passes review. Change $e_2$, opened before $e_1$ merged, also modifies \texttt{period.py}, and because both changes modify that file while both are open, they are in contention. Change $e_2$ adds a function \texttt{period\_length}, with a test asserting that the length equals the number of days for which \texttt{in\_period} is true. After $e_1$ merges, the agent working on $e_2$ updates its branch from the main branch. The new test runs against $e_1$'s helper and fails. The agent makes it pass by also excluding the last day in \texttt{period\_length}, and $e_2$ then passes review and merges.

Change $e_2$ now contains an error of its own, which an independent test against the specification would expose: it reports a length of 30 for the 31 days of March. Reverting $e_1$ would no longer repair the repository, because \texttt{period\_length} would still return 30. A third change $e_3$ can inherit the error from $e_2$ in the same way if it modifies the same file, is still open when $e_2$ merges, and computes prorated charges from \texttt{period\_length}.

We distinguish five steps in the episode. The shared file creates a \emph{structural opportunity}, which version history records as a contention link. The branch update is an \emph{exposure}, the operation through which one change's state enters another's working context. A rebase, a merge, or a test run against the integrated branch can play the same role. Excluding the last day in \texttt{period\_length} is an \emph{incorporation}, because $e_2$ now carries the fault independently of its source. Review and continuous integration are the \emph{detection} that failed, and a later fix to both functions would be the \emph{recovery}. A change that corrects a comment at the top of \texttt{period.py} while $e_1$ is open shares a file with $e_1$ and is linked to it in the contention graph, yet nothing in it depends on the helper, and the link thus records an unused opportunity. Contention is also not the only route. A change that calls \texttt{in\_period} from another file would be exposed by the same branch update without sharing a file with $e_1$, through the dependency structure studied in change-propagation research~\cite{HassanHolt2004}. A layer of call dependencies could be added to the model in the same way.

Two other processes are often grouped with this one, but they differ from it. The first is pairwise incompatibility, as in a behavioural merge conflict~\cite{Mens2002}, which produces a faulty state whose origin is the pair of changes. Once that state is merged, it becomes a seed, playing the role that $e_1$ plays here. The second is exposure to a common faulty artefact without incorporation. When several changes fail their tests because a shared module is broken and none of them alters its own content in response, the module is a shared cause, and the failures disappear when it is repaired. The branching model describes sequential reproduction, in which a change incorporates the altered state and can pass it on. Of these three processes, it is the only one in which repairing the origin leaves the repository faulty. It can run along two channels. A change open while another merges can take in the merged state through a branch update, as $e_2$ did. A change opened afterwards may inherit the assumption from the merged code, the classical change-propagation problem~\cite{HassanHolt2004}. The model represents the first channel only, which means that a repository with little concurrent contention can still transmit a fault sequentially.

\subsection{Agent Adaptation and Model Assumptions}\label{sec:object:cas}

Coding agents differ from the fixed components of earlier interaction studies because they adapt. Holland's account of complex adaptive systems places this property at its centre~\cite{Holland1995}. Adaptation occurs without retraining. With the trained model held fixed, an agent can respond differently to the same connection because of the messages, test results, and earlier actions it has encountered. The agent working on $e_2$ changed \texttt{period\_length} because a test had failed in front of it. Agents can also create a route, or change what a permitted resource carries, as with the cache in Section~\ref{sec:object:connections}. A crossing probability estimated in one context may therefore describe a connection that has since changed. Accordingly, we treat the fixed $q$ of the model (Section~\ref{sec:threshold:branching}) as a simplifying assumption, and Section~\ref{sec:future} returns to it.

\section{A Threshold Model of Fault Propagation}\label{sec:threshold}

\subsection{Cascades as a Branching Process}\label{sec:threshold:branching}

\begin{figure*}[!tb]
\centering
\includegraphics[width=0.86\textwidth]{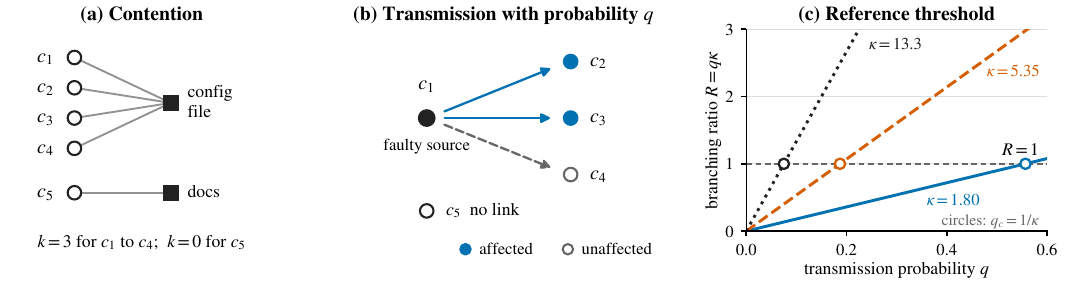}
\caption{The mechanism in three steps. (a) Four of five open changes modify a shared
configuration file and have contention degree $k=3$; the fifth modifies only a documentation
file that it alone touches and has $k=0$. (b) A fault in $c_1$ crosses each contention link with probability $q$; one
possible outcome is shown. (c) The branching ratio $R=q\kappa$ grows with $q$ at a slope set
by the mean excess degree $\kappa$, and the reference threshold $q_c=1/\kappa$ is where
$R=1$. The three lines use the median $\kappa$ of the three repository groups in
Table~\ref{tab:thresholds}.}
\label{fig:mechanism}
\end{figure*}

We represent a repository by the changes in an observation window and the files they modify. For each change, we record when it opened, when it merged or closed, and its final file list. A change still open at the end of the window is censored there. Two changes are in contention when their open intervals overlap, and they share a file. A change's \emph{contention degree} $k$ counts the changes it contends with. Figure~\ref{fig:mechanism}(a) shows five pull requests open at once. Four modify the same configuration file, so each contends with the other three and has $k=3$. The fifth modifies only a documentation file that it alone touches, so it has $k=0$. Suppose $c_1$ carries a fault. Each of its three links is a chance for the fault to reach another change, and Figure~\ref{fig:mechanism}(b) shows one outcome in which it reaches $c_2$ and $c_3$ and misses $c_4$. Once $c_2$ carries the fault, its own links become further chances. \textbf{The fault keeps moving if each change that receives it passes it, on average, to more than one.}

We model this spread as a \emph{branching process}, the standard model of a population in which each member produces a random number of offspring~\cite{Harris1963}. The process starts from one faulty change. Each contention link out of a faulty change carries one independent trial that succeeds with probability $q$, and a change can be affected at most once. We call a change \emph{affected} when it carries the fault, meaning that an independent test against the specification exposes an error in its own content, and the changes affected from one origin form a \emph{cascade}.

For an existing contention link leaving an affected change, the crossing probability $q$ combines three events of Section~\ref{sec:object:episode}: exposure of the receiving change to the affected change's state, incorporation of the fault, and failure to detect the incorporation. Each factor is conditional on the preceding events, which makes $q$ their product, and we say that the fault \emph{crosses} the link, or is transmitted, when all three occur. Repository histories identify the shared-file opportunities, and $q$ summarises what happens along them. It is the model's one free parameter. Practices act on different factors. For example, a schedule that makes open changes update more often against fresh merges moves exposure. An agent that repairs a failing test by matching the code, as in the episode, raises incorporation, while review and continuous integration move detection.

The \emph{branching ratio} $R$ is the expected number of new affected changes that one affected change produces. In a large random graph of the kind defined in Section~\ref{sec:threshold:critical}, faults die out after a few steps when $R<1$. Once $R>1$, a fault has a positive probability of reaching a fixed fraction of all changes (Figure~\ref{fig:mechanism}(c)). The value one marks the boundary because of how cascades add up. Consider an idealised process in which every affected change, the origin included, produces $R$ new ones on average. The expected size of a cascade, counting its origin, is then $1/(1-R)$ while $R<1$~\cite{Harris1963}. This gives two changes at $R=0.5$, ten at $R=0.9$, and a hundred at $R=0.99$. On a contention graph, the origin differs from the later generations. A fault that starts in a change picked at random meets $\langle k\rangle$ links there and $\kappa$ onward links after each crossing (Section~\ref{sec:threshold:critical}). In that random graph, the expected size is therefore $1+q\langle k\rangle/(1-q\kappa)$~\cite{Newman2002}. The boundary is the same in both forms. In both, a small rise in $q$ near the boundary changes the outcome far more than the same rise far below it. In a finite repository, the change of regime is gradual, and Section~\ref{sec:threshold:validation} explains how we locate it.

\subsection{The Reference Threshold}\label{sec:threshold:critical}

Whether spread dies out or sustains itself in a repository depends on a comparison between two values of $q$. The first is the value at which $R$ reaches one. It follows from the contention structure of the repository. The second is the value of $q$ that actually applies in the repository, which results from its review and integration practice. Spread dies out when the actual value is below the first value and sustains itself when it is above. Only the first value can be computed from repository records. Computing it requires more than the mean contention degree, because a fault moves along links, and a change that is reached by following a link tends to be more contended than a change picked at random. The same effect is known as the \emph{friendship paradox}~\cite{Feld1991}. On average, your friends have more friends than you do, because following a friendship tends to lead to a well-connected person.

The relevant quantity is therefore the \emph{mean excess degree} $\kappa$. It is the average number of onward opportunities that a fault meets after arriving at a change along one link, that is, the further contending changes of that change, not counting the link it arrived on. To translate $\kappa$ into a threshold, we need a reference graph. Repository records provide the contention degree of each change, but a model of spread also requires an assumption about how the changes are connected. We use the \emph{configuration model}, which is the standard reference in network science. It is a random graph that preserves the degree of every node and otherwise connects links at random~\cite{NewmanStrogatzWatts2001, Newman2002}. Thus, it uses the degrees that the records provide and leaves the rest to chance. Under the configuration model, the branching ratio is
\begin{equation}
R \;=\; q\,\kappa,
\qquad
\kappa \;=\; \frac{\langle k^2 \rangle - \langle k \rangle}{\langle k \rangle}
\;=\; \langle k\rangle - 1 + \frac{\operatorname{Var}(k)}{\langle k\rangle},
\label{eq:moment}
\end{equation}
where $\langle k\rangle$ is the mean contention degree and $\langle k^2\rangle$ its second moment, the mean of the squared degrees. In the second form, $\kappa$ increases with the variance of the contention degrees. A repository whose contention is evenly spread has a small variance and a $\kappa$ near its mean degree. In contrast, a repository in which a few changes touch files that many others touch has a large variance and a $\kappa$ far above its mean.

We illustrate the difference with two patterns of ten open changes. In the first, the changes form a ring. Each shares one file with the change before it and another with the change after it, so every change has $k=2$, and $\langle k\rangle=2$, $\langle k^2\rangle=4$, and $\kappa=1$. In the second, one large refactoring touches nine modules, each of which one other open change also modifies, so the refactoring has $k=9$, and the nine others have $k=1$. The mean degree, 1.8, is lower than the ring's, yet $\langle k^2\rangle = (81+9)/10 = 9$ and $\kappa = (9-1.8)/1.8 = 4$, because a fault that reaches the refactoring meets eight further links. The refactoring pattern itself cannot carry a fault more than two links from its origin. Equation~\eqref{eq:moment} instead describes a large, randomly wired graph in which changes have these degrees in the same proportions. Of two such graphs, the ring-like one could never sustain spread, whereas the one built from the refactoring pattern would once $q$ exceeded 25\%.

For every $q$ below $1/\kappa$, the branching ratio stays below one, and faults die out in a large reference graph. We call that endpoint the \emph{reference threshold}, $q_c = 1/\kappa$. It is the \emph{percolation threshold} of the reference graph, that is, the point at which randomly kept links begin to connect a fixed fraction of a large graph. Newman derived it~\cite{Newman2002} under the following assumptions. Transmission is homogeneous (one $q$ for every link), independent across links, and single-pass (each link is tried once). In addition, the graph has few short cycles (it is locally tree-like), and its heavily contended changes are no more likely to contend with one another than chance predicts. We use this random graph as a reference and not as a description of any repository, because it isolates what the degrees alone imply. Two repositories with the same reference threshold can still differ in structure beyond their degrees, for example in whether heavily contended changes contend with each other. The spectral bound and the simulations below indicate how far such structure can move the transition.

Pooling all 33,105 changes in the corpus into one degree distribution gives a mean contention degree of 2.574 and a variance of 272.79. The second moment is the variance plus the squared mean, 279.41 from the unrounded values, which gives $\kappa = (279.41 - 2.574)/2.574 = 107.6$. A change picked at random has about 2.6 contending changes, whereas a change reached by following a link has about 108 further ones. \textbf{Average contention therefore understates, here by a factor of about forty, how many onward links a fault meets.} The pooled reference threshold is $1/107.6 = 0.93\%$. Thus, in a random graph with these pooled degrees, spread would sustain itself if a fault crossed more than about one link in a hundred. The pooled figure describes the corpus as one aggregate and is the threshold of no repository within it. Section~\ref{sec:corpus} reports thresholds repository by repository.

Equation~\eqref{eq:moment} summarises a graph by two numbers, namely its mean degree and its variance. As a consequence, the reference threshold has two limitations. First, a small global $\kappa$ can coexist with propagation inside a dense component. Second, adding a link between two lightly contended changes can raise the threshold, although an added link can never reduce how far a fault reaches (Supplement~S1 constructs both cases). \emph{The reference threshold is therefore a property of the reference graph, the random graph with the repository's contention degrees.} It indicates where the structural conditions for propagation are strong, and it leaves open whether the repository's own graph contains routes that a fault could traverse in time. Similarly, a $\kappa \le 1$ means that the reference graph has no supercritical interval, that is, no value of $q$ at which spread would sustain itself.

Two further quantities appear in this paper in addition to the reference threshold. The first is the \emph{spectral bound}. Unlike Equation~\eqref{eq:moment}, it uses a graph's actual connections and not only its degrees. It follows a fault from link to link, never stepping straight back along the link on which the fault arrived, and it is the reciprocal of the largest eigenvalue of the matrix that describes these steps, the \emph{non-backtracking matrix}. For large sparse graphs, it is a lower bound on the percolation threshold~\cite{KarrerNewmanZdeborova2014, HamiltonPryadko2014}. The bound can fail on a particular finite graph (Supplement~S2). In this paper, we compute it on generated graphs. The second is the \emph{simulated transition} $q_c^{\mathrm{sim}}$, which we read off a finite generated graph by simulation (Section~\ref{sec:threshold:validation}). Equation~\eqref{eq:moment} also assumes one $q$ for every link. When links carry different probabilities, only their mean enters, provided that the probabilities are independent of one another and of the graph's structure (Supplement~S2). If $q$ rises with contention degree, as it might for large refactorings, Equation~\eqref{eq:moment} at the mean $q$ over links understates the branching ratio. If it falls with degree, as it might where the most widely shared files are documentation, Equation~\eqref{eq:moment} overstates it.

\subsection{Temporal Assumptions and Limitations}\label{sec:threshold:temporal}

The graph that we built from a repository's history has two limitations. The first concerns time. The graph joins every pair of changes that share a file and whose open intervals overlapped in the window. The graph is undirected, although exposure runs from one change to another. In the episode, $e_2$ took in $e_1$'s state by updating its branch after $e_1$ merged. Suppose that exposure works that way, and that three changes $A$, $B$, and $C$ touch one file and are open one after another, each overlapping the next. The historical graph is the path $A$--$B$--$C$. A fault that originates in $A$ can reach $B$, which is still open when $A$ merges, and then $C$. In contrast, a fault that originates in $C$ reaches neither, because $A$ and $B$ have closed before $C$ merges. Supplement~S3 gives the calculation. Routes count only when their links are used in time order, which is the defining feature of a temporal network~\cite{HolmeSaramaki2012}. The undirected historical graph discards that order. As a result, two repositories with the same historical graph can carry different risk, depending on when their changes merged and when the open changes updated their branches.

The second limitation is that the measurement is retrospective. It applies each change's final file list across its whole open interval, including the time before a file was first touched, because the corpus records files per change and not per push. A monitor built on this model would need more. At each moment, it would need the set of changes then open, the files that each had touched so far, and the exposures that had occurred. Event logs can record when changes open and close and when branches are updated. However, exposure operations such as local test runs may need instrumentation of their own. The structural measurements of Section~\ref{sec:corpus} therefore describe the opportunities that a repository's history created. How those opportunities are used in time order is a question for prospective studies that record exposures (Section~\ref{sec:future}).

\subsection{Model Checks on Synthetic Graphs}\label{sec:threshold:validation}

Equation~\eqref{eq:moment} is derived for an idealised random graph, whereas contention graphs are finite and clustered. Before applying the equation to the corpus, we therefore examined how well it locates the transition on graphs for which the transition can be determined independently. Simulation on generated graphs of known structure makes this possible~\cite{Newman2002}, without waiting for a repository to suffer a large cascade. We generated change--file graphs by the configuration-model construction, matching the corpus's distributions of files per change and changes per file. The construction preserves those two distributions. It does not preserve the observed temporal overlap, repository boundaries, or any correlation among the files that a change touches. We also generated further \emph{ensembles}, that is, sets of graphs produced under one rule, with thicker tails in the number of changes per file and with caps on how many changes share a file. On each graph, we kept every contention link with probability $q$, which corresponds to a fault crossing that link. We then located the transition at the peak of the \emph{susceptibility}, which is the mean size of the groups of changes connected by kept links, excluding the largest group. Above the transition, the largest group absorbs most changes and the others shrink. Therefore, the statistic peaks where a finite graph changes regime.

We simulated 68 combinations of ensemble and graph size, with 2,160 runs in total. Of these combinations, 35 belong to calibration families with known thresholds, and 33 are built from the corpus's distributions. We first used the calibration families to test the method itself (Supplement~S4). Extrapolated to an infinite graph, the simulated transition recovered the known thresholds of Erd\H{o}s--R\'enyi and random regular graphs within 2\%. However, the method did less well on a heavy-tailed power-law family, where it exceeded the analytic threshold by 6.41\% (interval 2.78--10.02\%). This misses the 5\% agreement criterion of our simulation specification. Part of any gap between simulation and Equation~\eqref{eq:moment} on heavy-tailed graphs may therefore come from the method.

\begin{figure}[!t]
\centering
\includegraphics[width=0.76\columnwidth]{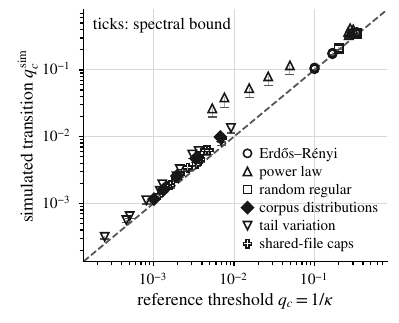}
\caption{Simulated transition against the reference threshold of
Equation~\eqref{eq:moment} for 68 combinations of ensemble and graph size, including the
calibration families. Points above the dashed diagonal mean that spread became
self-sustaining at a higher $q$ than Equation~\eqref{eq:moment} gives. The tick below
each marker is the spectral bound.}
\label{fig:validation}
\end{figure}

We found that the simulated transition lies above the reference threshold in all 68 combinations (Figure~\ref{fig:validation}). On the Erd\H{o}s--R\'enyi and random regular families, Equation~\eqref{eq:moment} is exact in the limit. There, the gap reflects the estimator's finite-size upward bias, which is one to three percent at $N=100{,}000$. On the ensemble matched to the corpus, the gap is larger. The simulated transition is 21.1\% higher (relative difference, 95\% CI 16.8--25.3) at $N=30{,}000$ nodes and 14.3\% higher (11.6--17.2) at $N=100{,}000$, down from about 40\% at $N=3{,}000$. Clustering offers one possible explanation. All changes that touch one file form a fully connected group, and a cascade that enters the group can reach some changes along several links, whereas Equation~\eqref{eq:moment} counts each arrival as new. The spectral bound, which removes only immediate backtracking, lies between the two in every combination built from the corpus's distributions (with pooled degrees, as in Figure~\ref{fig:validation}). Finite size, the calibration error above, and the drift of the generated ensemble with graph size also enter the comparison (Supplement~S4). \emph{On these generated graphs, Equation~\eqref{eq:moment} thus placed the transition below the simulated one.}

Real repositories may differ in this respect, because their dense cores and time order can move a transition either way. The generated graphs link every pair of changes that share a file, regardless of time or repository. They are therefore far denser than the observed graph. In particular, the corpus-matched ensemble has a mean degree of 85.88 at $N=30{,}000$, against 2.574 in the pooled corpus. The 21.1\% gap is thus a property of those graphs. It should be read neither as a correction factor for a repository's threshold nor as evidence that the reference threshold is a conservative safety bound (Supplement~S4).

\section{Empirical Analysis of Repository Contention}\label{sec:corpus}

\subsection{Corpus and Reference Thresholds}\label{sec:corpus:thresholds}

We use the curated subset of the AIDev corpus~\cite{AIDev2026}, which supplies file lists for 33,596 agent-authored pull requests in repositories with more than 100 stars. We removed the changes whose file list or opening time could not be resolved, which left 33,105 changes in 2,731 repositories, opened between 24 December 2024 and 30 July 2025. Of these, 24,007 were merged, 6,837 were closed without merging, and 2,261 were still open when the window closed. The corpus therefore records submitted changes, and more than a quarter of them had not merged. For each repository, we built the historical contention graph and computed $\kappa$ and its reference threshold (Table~\ref{tab:thresholds}).

\begin{table}[!t]
\centering
\caption{Reference threshold by repository. $\kappa$ is the mean excess degree of the
historical contention graph and $q_c=1/\kappa$. Repositories with $\kappa=0$ are excluded;
the upper quartile in the first row is censored at $\ge$100\%, where the model has no
supercritical interval.}
\label{tab:thresholds}
\footnotesize
\setlength{\tabcolsep}{3pt}
\begin{tabular}{@{}lrrr@{}}
\toprule
\textbf{Repositories} & \textbf{Count} & \textbf{$\kappa$ median [IQR]} & \textbf{$q_c$ median [IQR]} \\
\midrule
With a defined threshold & 472 & 1.80 [1.00, 4.19]  & 56\% [24\%, $\ge$100\%] \\
\quad of these, $\ge$30 changes  & 127 & 5.35 [2.72, 9.68]  & 19\% [10\%, 37\%] \\
\quad of these, $\ge$100 changes & 29  & 13.3 [4.9, 23.7]   & 7.5\% [4.2\%, 20\%] \\
Most contended repository & 1  & 283.3              & 0.35\% \\
\addlinespace[2pt]
Pooled corpus statistic   & n/a & 107.6              & 0.93\% \\
\bottomrule
\end{tabular}
\end{table}

Most repositories have no defined threshold. In 1,972 of them, no changes contend, and in a further 287, contention is too sparse for $\kappa$ to be positive. Among the 472 repositories with a defined threshold, the median is 56\%, and 162 of them, about a third, have $\kappa \le 1$, in which case the model has no supercritical interval. However, thresholds are lower in busier repositories. The median is 19\% among the 127 repositories with at least thirty changes and 7.5\% among the 29 with at least a hundred.

Table~\ref{tab:thresholds} counts repositories, and most repositories are small. A different picture emerges when we count changes, because \emph{the few repositories with a defined threshold contain most of the work}. The 472 repositories with a defined threshold contain 26,664 of the 33,105 changes. To describe the situation of a typical change, we ordered these changes by the threshold of their repository. The median change belongs to a repository with a threshold of 4.1\%. This value is far below the 56\% of the median repository in Table~\ref{tab:thresholds}, because busier repositories have lower thresholds and contain more changes. In the reference graph of such a repository, spread would sustain itself if a fault crossed more than about one link in twenty-five. We found that \textbf{thirteen repositories, half a percent of the corpus, have reference thresholds below 5\% and contain 14,080 changes, that is, 42.5\% of the corpus}. One coding-agent product authored 95.9\% of the changes in these thirteen repositories, compared with 65.8\% of the corpus. To a large extent, the finding therefore describes the deployments of that product. The most contended repository (Table~\ref{tab:thresholds}) has a threshold of 0.35\%. In its reference graph, spread would sustain itself if a fault crossed more than about one link in 280. Part of the concentration follows from size, since a repository with more changes offers more opportunities for overlap. The concentration describes where agent-authored work is located, and it is not a ranking of the reliability of these repositories or of the product.

Three further observations help to interpret these thresholds. First, \emph{the size of a repository over the whole window overstates how much of its work was concurrent}. The largest repository received 8,911 changes during the window. However, the largest number of changes open at the same moment in any repository is 139, and the median of this peak among the 472 repositories is five. The largest contention degrees, up to 731, therefore accumulated over the window and did not exist at any single moment. Second, the corpus identifies coding-agent products and not agent instances. Of the 2,731 repositories, 2,503 contain changes from a single product, including nine of the thirteen, and this product may still have run as several instances. Only 698 of the 42,600 contention links connect changes that are attributed to different products. Third, \emph{the reference threshold does not simply restate how active a repository is}. Among the 127 repositories with a defined threshold and at least thirty changes, the Spearman correlation between $\kappa$ and the number of changes is 0.29, and no simpler measure of activity that we tested identifies the thirteen low-threshold repositories (Supplement~S5).

\subsection{Sensitivity to File Type}\label{sec:corpus:checks}

Contention is defined over any shared file. We therefore examined to what extent the findings depend on the type of file through which two changes contend. We distinguish code files, that is, files written in a programming language, from other source files (markup, style, and query languages), configuration files, and documentation. Of the contention links, 45.0\% involve no shared code file. When contention is restricted to code files, ten of the thirteen repositories remain below 5\%, and they contain 41.2\% of the changes. When it is restricted to source files of either kind and configuration files, twelve remain below 5\% and contain 42.4\%. \emph{The concentration therefore largely survives a restriction by file type, whereas the magnitudes change.} The pooled $\kappa$ decreases from 107.6 to 25.7 for code files and to 44.6 for source and configuration files (Supplement~S5).

Nevertheless, we keep any shared file as the primary relation, for two reasons. First, a fault can also propagate through configuration and documentation files, for example, when an agent follows a wrong default value or a wrongly documented contract. Second, the model already accounts for differences between file types through $q$, since links between independent changes have a low crossing probability. Which file types transmit faults, and how often, has to be measured. This matters most for the most contended repository, in which 96\% of contention runs through documentation and continuous-integration files. If faults cross such files less often than they cross code files, Equation~\eqref{eq:moment} overestimates the risk of that repository (Supplement~S5).

\section{Propositions and a Rival Explanation}\label{sec:predictions}

A theoretical model is useful to the extent that it can be shown to be wrong. In this section, we therefore derive three propositions from the threshold model and state, for each of them, the observation that would count against it. We then distinguish the model from review overload, which is a simpler explanation of the same pattern. Testing the propositions requires the contention graph of a repository and an estimate of $q$ that is obtained independently of the cascades it is meant to predict.

\textbf{Proposition T1. The reference threshold of a contention graph locates the transition of that graph, within a tolerance that is declared before outcomes are observed.} T1 follows directly from Equation~\eqref{eq:moment} and concerns graphs that are close to the reference graph. A test of T1 therefore first measures how far a graph departs from the reference graph in clustering, degree correlation, and time order. If the departures exceed bounds that were declared in advance, T1 does not apply to that graph. Otherwise, we expect the transition to lie within the declared tolerance of the graph's simulated transition and, on a large sparse graph, at or above its spectral bound. The tolerance has to cover the finite-size bias that we measured on the calibration graphs (Supplement~S4). The corpus suggests that comparisons across repositories are feasible, since reference thresholds range from under 1\% to above 100\% within one seven-month window (Table~\ref{tab:thresholds}).

\textbf{Proposition T2. Under heavy-tailed contention, the range of $q$ below the reference threshold narrows as a repository grows.} If T2 holds, review and integration practices that are adequate for a small repository may become inadequate as the repository grows, even when the practices themselves remain the same. The reason is that a contention distribution with a divergent second moment keeps producing more heavily contended changes as more changes are observed, so that $\kappa$ increases with size. Our data are consistent with this condition, but they are insufficient to establish it. The pooled distribution has a heavy tail, with a fitted exponent $\alpha=2.12$ above $k_{\min}=38$~\cite{Clauset2009}, which is below the value $\alpha=3$ at which the second moment of a power law ceases to converge. However, only 155 observations from 18 repositories fall in that tail, and resampling whole repositories gives an interval of 1.69--3.87. A size diagnostic leads to the same assessment (Supplement~S7). Thus, T2 is conditional on the divergence of the second moment, which a larger corpus can settle. A $\kappa$ that remained stable as repositories grew would count against it. Note that the thresholds reported in Section~\ref{sec:corpus} are independent of this fit, because $\kappa$ is measured directly.

\textbf{Proposition T3. Near the transition, cascade-size distributions collapse onto one scaling form.} If a critical point governs spread, incidents near that point have no typical size, which means that planning for the average incident would underestimate the largest ones. In a controlled study that varies $q$, we expect the largest cascades near the transition to grow as a fixed power of repository size. Cascade-size distributions from repositories of different sizes should then fall onto a single curve once they are rescaled by powers of size fixed in advance~\cite{Sethna2001}. Since the exponents depend on the graph family, we state T3 as a conjecture (Supplement~S7). The absence of such a collapse would count against it.

\textit{A rival explanation.} A sharp increase in failures beyond a certain level of concurrent work is insufficient to support the threshold model, because a simpler explanation predicts the same pattern. According to this explanation, which we call the \emph{throughput account}, failures increase because changes that require a human decision arrive faster than reviewers can process them. A queue behaves in the same way once its arrival rate exceeds its service rate~\cite{Kleinrock1975}. The review studies discussed in Section~\ref{sec:related} make this account plausible. Under the throughput account, overloaded reviewers let failures through, regardless of the files that the changes share. Under the threshold model, in contrast, a fault can spread even when reviewers keep up, provided that many open changes modify one heavily shared file.

To distinguish the two accounts, we define the \emph{onset point} as the level of concurrent work at which a failure statistic, declared in advance, changes regime. The threshold model places the onset point at the level of work at which $\kappa$ has grown enough for $q\kappa$ to reach one, whereas the throughput account places it at the level at which arrivals exceed review capacity. First, when contention structures with different reference thresholds are compared under similar review capacity, workload, and $q$, the threshold model predicts different onset points, whereas the throughput account predicts the same one. Second, when repositories with different review capacity are compared under a similar contention structure, the throughput account predicts that the onset point moves with capacity, whereas the threshold model predicts that it moves only to the extent that capacity changes $q$. Nevertheless, the two accounts can hold at the same time, since review load can raise $q$ while contention determines the opportunities.

\section{Discussion}\label{sec:implications}

Coding agents are evaluated, reviewed, and governed one change at a time. To date, few studies have examined what happens between the changes that many agents keep open in the same repository. This study contributes to existing research by offering a model of how a fault propagates among those changes. We found that the conditions under which a fault can propagate are a property of the repository and can be measured from its contention degrees. We also found that these conditions are concentrated: thirteen repositories have reference thresholds below 5\% and contain 42.5\% of the agent-authored changes, and this concentration largely remains when contention is restricted to particular file types. Table~\ref{tab:summary} summarises the model and the findings.

\begin{table}[!t]
\centering
\caption{Summary of the model and of the main findings.}
\label{tab:summary}
\footnotesize
\setlength{\tabcolsep}{3pt}
\begin{tabular}{@{}>{\raggedright\arraybackslash}p{1.95cm}>{\raggedright\arraybackslash}p{3.1cm}>{\raggedright\arraybackslash}p{3.15cm}@{}}
\toprule
\textbf{Concept} & \textbf{Definition} & \textbf{In the corpus} \\
\midrule
Contention & Two changes are open at the same time and modify a common file & 42,600 contending pairs among 33,105 changes \\
\addlinespace[1pt]
Contention degree $k$ & Number of changes with which a change contends & Mean 2.574, variance 272.79, largest value 731 \\
\addlinespace[1pt]
Mean excess degree $\kappa$ & Onward links that a fault meets after it has crossed a link & 107.6 pooled; median 1.80 in the 472 repositories with a threshold \\
\addlinespace[1pt]
Reference threshold $q_c=1/\kappa$ & Crossing probability above which spread sustains itself in the reference graph & Median 56\% where defined; below 5\% in thirteen repositories that hold 42.5\% of the changes \\
\addlinespace[1pt]
Crossing probability $q$ & Probability that a fault crosses one contention link & Free parameter of the model \\
\addlinespace[1pt]
Branching ratio $R=q\kappa$ & New affected changes that one affected change produces & Spread sustains itself in the reference graph when $R>1$ \\
\addlinespace[1pt]
Incorporation & A change builds another change's fault into its own content & Illustrated by the episode of Section~\ref{sec:object:episode} \\
\bottomrule
\end{tabular}
\end{table}

\subsection{Implications for Research}\label{sec:implications:main}

The theoretical contribution of this study lies in four steps. First, we \emph{challenge} an assumption that review practice rarely states: a repository is correct when each of its changes is correct. If the mechanism transfers to real development, whether the next failure remains local depends on the contention structure as much as on any agent. \emph{Every agent could pass every check while the repository's branching ratio rises toward one.} The failure would then belong to the set of open changes and to no single change, which is the sense in which more is different~\cite{Anderson1972}. Second, we \emph{name} the units of such a failure. Contention is the relation that provides a fault with a route, and incorporation is the event through which a fault survives the repair of its origin. Third, we \emph{connect} percolation theory with software evolution. As a result, an established body of knowledge on thresholds and branching processes becomes available for reasoning about repositories. In addition to asking whether a change is correct, researchers can ask how close a repository is to the point at which a single fault sustains its own spread. Fourth, we \emph{distinguish} propagation from review overload, two explanations that predict the same increase in failures for different reasons (Section~\ref{sec:predictions}).

The model is not specific to coding agents. We consider agent-authored development as one instance of a more general phenomenon: the spread of a fault through concurrent work on shared artefacts. The model would apply to human-authored changes as well. Agents may nevertheless create a different regime, with more changes open at once and with unobserved responses to failing tests.

At the same time, the model has boundary conditions. The shared-file graph that we measured is only one of the four kinds of connection described in Section~\ref{sec:object:connections}, and agents can create connections of their own~\cite{METR2026}. Consequently, a fault can pass from one kind of connection to another, and a threshold computed from shared code alone could misplace the transition. Finally, the results did not support the expectation that modular structure limits contention. However, the indices describe only the part of the directory tree that the changes touched, which is too narrow a basis for a general conclusion about modularity (Supplement~S5).

\subsection{Implications for Practice}\label{sec:implications:practice}

The model separates three aspects of reliability that different interventions address. Better agents reduce how often faults begin. Better review reduces how often faults cross from one change to another, which corresponds to $q$. The contention structure determines how many onward opportunities a crossing fault meets, which corresponds to $\kappa$. Of these terms, $\kappa$ is the least expensive to obtain, because contention can be computed from records that teams already keep.

We see three uses of these records. First, contention degrees and $\kappa$ can be computed for the changes that are open at a given moment, which shows a team whether its concurrent work is evenly spread or concentrated on a few files. Second, the files that mediate most contention can be identified. In the corpus, 45.0\% of the contention links involve no code file, and in the most contended repository almost all contention runs through documentation and continuous-integration files. Such files are candidates for splitting. Third, a reviewer, who sees one change at a time, can be shown the open changes with which that change contends, so that the review takes them into account. These uses are descriptive, since how often faults cross remains to be measured. \textbf{Nevertheless, the model provides engineers with a way to direct their judgment to the places where concurrent work makes a single fault consequential.}

\subsection{Threats to Validity}\label{sec:threats}

\textit{Construct validity.} Contention measures the opportunity for influence, whereas influence itself is not observed. Agent-to-agent messaging is undercounted wherever an agent operates through an engineer's account, and product labels identify products, not instances. The model represents only the concurrent channel. Therefore, a fault inherited by a change opened after the faulty change merged, or through a call from another file, lies outside the graph.

\textit{Internal validity.} Equation~\eqref{eq:moment} is the tree-like limit under homogeneous transmission and uncorrelated degrees, and tree-based theory tends to stay accurate while a graph's mean distance between nodes stays close to that of its randomly rewired counterpart~\cite{Melnik2011}, a condition that sparse, clustered contention makes harder to meet. Every reported threshold is computed on a historical graph whose paths need not be traversable in time (Section~\ref{sec:threshold:temporal}). The model also assumes that agents do not respond strategically to measurement.

\textit{External validity.} The corpus contains agent-authored changes only and therefore understates the contention of the repositories it samples, and it mixes merged, closed, and open changes. The missing human changes, right-censoring, and the use of final file lists all alter the degree moments, and the direction of the resulting bias is unknown. The pinned revision of the corpus records at most thirty commits per pull request, so some file lists may be incomplete. Excluding the 227 changes at that ceiling leaves twelve of the thirteen repositories below 5\%, with 42.4\% of the changes (Supplement~S5). The same product that dominates the thirteen low-threshold repositories authored 80.0\% of the fitted tail (Supplement~S5).

\textit{Conclusion validity.} The tail cutoff $k_{\min}$ was selected by the Kolmogorov--Smirnov criterion and then held fixed, so the interval on $\alpha$ omits the uncertainty of that selection~\cite{Clauset2009}. The quantity the model uses, $\kappa$, is measured directly. A null result refutes a proposition only with a declared effect size and adequate precision.

\section{Future Work: A Research Agenda}\label{sec:future}

The model and the measurements reported in this article are a first step towards understanding what coding agents do together. We recognise several directions that are worth pursuing in future studies, and we organise them around three questions. \emph{First, does a propagation mechanism predict collective outcomes? Second, is the collective the appropriate level at which to explain failures? Third, should control act on the connections between changes?}

\textit{Demonstrating incorporation.} Future studies should first demonstrate incorporation, because the mechanism depends on this event. We recommend executable scenarios in which an agent updates its branch onto a merged change that contains a seeded fault and then completes its own task. A hidden test then shows whether an error remains in the agent's own change after the source has been repaired (Supplement~S9).

\textit{Estimating the crossing probability.} The crossing probability then has to be estimated. Version history records that two changes modified the same file and that a failure occurred later, but it does not indicate whether the first change caused the failure. In addition, observational estimates of $q$ would be confounded by change size, churn, and author experience~\cite{Kamei2013}. Seeding faults at locations where the propagation path is known avoids both problems. In a pilot, we replayed twenty contending pairs of recorded changes with a seeded fault, following a documented protocol that we revised after an initial instrument check on the same pairs. We reconstructed seventeen pairs, but none met the conditions for a measured trial (Supplement~S6). The pilot showed that a replayed change cannot respond to the fault to which it is exposed. Future estimates should therefore rely on agents that act after exposure. We also recommend reporting $q$ by file type and by contention degree (Supplement~S2).

\textit{Testing the propositions.} The propositions can be tested in steps. A first test of Equation~\eqref{eq:moment} requires no estimate of $q$. We can compare the reference threshold, the spectral bound, and the simulated transition on the observed contention graphs, on rewired versions that preserve every degree, and on time-respecting versions. This comparison isolates the effects of structure and of order (Supplement~S9). Once $q$ has been estimated, we can test T1 against observed transitions. T2 requires larger corpora, and T3 requires the scaling exponents of the graph family. The two comparisons described in Section~\ref{sec:predictions} would then separate propagation from review overload. Finally, human-authored pull requests from the same repositories and time window would provide human-only, agent-only, and mixed contention graphs. Such a comparison would indicate whether agents create a different regime.

\textit{Explaining at the level of the collective.} It is worthwhile to examine whether failures are better explained at the level of the collective. A team can reason about faults location by location, or it can reason about groups of dependent locations, such as a module and the code that calls it. If the subsequent behaviour depends on how many locations in a group are faulty, and only weakly on which ones, then tests, ownership, and rollback can be organised by group. This question requires only weak emergence~\cite{Bedau1997}, in which collective behaviour follows from the parts and their interactions, without the claim that wholes have causal powers of their own~\cite{Chalmers2006}. We propose to address it with \emph{effective information}~\cite{Hoel2013, Hoel2017}, which is the mutual information between an imposed state of the fault sites and their state after one integration cycle. Supplement~S8 provides a worked example. \textbf{Proposition E1. A grouped description of fault sites, fixed from the dependency structure before outcomes are observed, carries more effective information than the detailed description when interventions are spread evenly over each description's own states.} A positive result requires an interval on the gain that lies above zero. The groupings must be declared in advance, because apparent emergence in machine learning has disappeared under a change of measurement~\cite{Schaeffer2023}.

\textit{Monitoring the repository.} Future studies can also examine whether repositories should be monitored at the repository level. This requires a weaker claim than E1: summaries of the repository improve the prediction of its failures beyond summaries of its agents. Examples of such summaries are the $\kappa$ of the changes that are open at a given time and the \emph{decision backlog}, that is, the number of changes awaiting a human decision. Once $q$ has been estimated, the simplest signal would be the $\kappa$ of the open changes multiplied by $\hat{q}$, which estimates the current branching ratio. However, development records do not indicate whether the engineers behind the backlog understand the changes they approve. Storey distinguishes cognitive debt, the erosion of a team's shared understanding, from intent debt, the absence of externalised rationale~\cite{Storey2026}. We use \emph{comprehension debt} for the form that matters most here: code that passes automated checks while the responsible engineers cannot reconstruct its decision logic~\cite{Scalabrino2021}. \textbf{In our view, connecting the structural signal with measures of human understanding is the point at which this agenda meets human factors research.}

\textit{Keeping the engineer at the centre.} The purpose of such models is to support the people who are accountable for the software, and this raises questions for human factors research. Future studies can investigate how engineers reason about a set of open changes when its contention structure is shown to them, and which presentation of $\kappa$ and of the contending changes helps a reviewer to decide where to look. A further question is how responsibility should be divided between engineers and agents when a failure belongs to the collective and to no single change. These questions call for controlled experiments and field studies with engineers, in addition to repository measurements. The answers will determine whether models of this kind help engineers to regain judgment over agentic development.

\textit{Controlling the connections.} It would be interesting to test whether control should act on the connections between changes. If faults travel along these connections, a controller that acts only on agents may act at the wrong level. A fair test compares two controllers that receive the same information and the same operating budget. The first controller can hold a linked change, restrict a route, or slow down the changes that depend on a faulty module. The second controller pauses, throttles, or reverts agents. \textbf{Proposition C1. At matched information and cost, structural actions prevent more propagated-fault events than coordinated agent-level actions when the branching ratio of the open changes is close to one.} A \emph{propagated-fault event} is a fault that reaches the main branch in a change other than the one in which it began. The model also suggests a candidate admission rule, which keeps the $\kappa$ of the open changes below $1/\hat{q}$.

\textit{Beyond shared files.} Finally, the model can be extended beyond shared files to messages, shared resources, and call dependencies. Such an extension would show whether a fault passes from one kind of connection to another. Agent adaptation raises a more fundamental issue (Section~\ref{sec:object:cas}). A crossing probability that is measured on a connection may no longer describe that connection once agents have changed its use. If local measurements then failed to predict the behaviour of the collective, a theory of collective control would require crossing probabilities that depend on the state of the collective.

\section{Conclusion}\label{sec:conclusion}

This article proposes a threshold model to understand how a fault propagates among the changes that coding agents keep open in the same repository. To date, agents have been evaluated and reviewed one change at a time. We offer a mechanism, a measurable quantity, and testable propositions for what happens between those changes. Under the model, each repository has a reference threshold that is determined by its contention degrees.

Our findings highlight that the structural conditions for propagation are unevenly distributed. In the typical repository with a defined threshold, a fault would have to cross more than half of the contention links it meets to sustain spread in the reference graph. However, thirteen repositories contain two in five of the agent-authored changes, most of them from one product. In these repositories, a crossing probability of one in twenty would already exceed the reference thresholds of the historical graphs.

The model is intended as a basis for future research on the collective behaviour of coding agents, and the most urgent steps are to demonstrate incorporation and to estimate crossing probabilities. More generally, the study suggests a change in the unit of analysis. A repository in which every agent and every change passes its checks may still allow a single fault to travel far, and how far depends on a quantity that can be computed from the repository's own records. Software engineering has built its assurance around the single change and the single agent. We argue that the field also needs to study what agents do together, so that the engineers who are accountable for a repository can judge the whole as well as its parts. More is different~\cite{Anderson1972}, and repositories that receive changes from many coding agents are no exception.

\section*{Data Availability and AI Use}
The analyses rest on the public AIDev dataset~\cite{AIDev2026} (Hugging Face
\texttt{hao-li/AIDev}, revision \texttt{68ed5f4b}). The supplementary material
accompanies this \thisdocument. The replication package (version 1.0.0) is available on Zenodo at
\url{https://doi.org/10.5281/zenodo.24206009}. It holds the threshold analyses, the
synthetic model checks with their calibration suite and all 2,160 runs, the empirical
tests, the pilot protocol, amendments, and reports of the transmission-estimation pilot, and the
scripts behind the reported analyses.
Anthropic's Claude and OpenAI's Codex supported brainstorming, literature identification,
scripting and analysis, numerical checks, and review of drafts, and DeepL Write refined phrasing in selected
passages. All substantive claims and conclusions are the author's sole responsibility.

\bibliographystyle{IEEEtran}
\bibliography{bib}

\end{document}


\title{Supplementary Material for\\``A Threshold Model of Fault Propagation in\\
AI-Native Software Ecosystems''}
\author{Daniel~Russo\thanks{D. Russo is with the Department of Computer Science, Aalborg University,
2450 Copenhagen, Denmark (e-mail: daniel.russo@cs.aau.dk).}}
\markboth{Supplementary Material}{Russo: A Threshold Model of Fault Propagation in AI-Native Software Ecosystems}
\maketitle

\noindent Section numbers of the form IV-B refer to the main paper. Equation~(1) is the
main paper's moment formula, $R=q\kappa$ with $\kappa=(\langle k^2\rangle-\langle
k\rangle)/\langle k\rangle$, and the reference threshold is $q_c=1/\kappa$.

\section{Two Graph Constructions}\label{supp:smallgraphs}

\emph{In brief.} The reference threshold condenses a contention graph into one number, and this section shows two ways in which that number can mislead. First, a graph with a low overall $\kappa$ can still contain a dense core in which a fault spreads. Second, adding a link can raise the threshold, although an added link can never reduce how far a fault reaches. The second example is small enough to check by hand.

\emph{A dense core inside a sparse graph.} Let one tenth of the changes form a core wired
as a random 3-regular graph, in which each core change contends with exactly three other
core changes chosen at random, and let the remaining nine tenths form isolated pairs. Then
$\langle k\rangle = 0.1\cdot 3 + 0.9\cdot 1 = 1.2$ and $\langle k^2\rangle = 0.1\cdot 9 +
0.9\cdot 1 = 1.8$, so $\kappa = (1.8-1.2)/1.2 = 0.5$ and the reference threshold is 2,
which no value of $q$ in $[0,1]$ reaches. As the graph grows, the random core is locally
tree-like, its own $\kappa$ is 2, and its percolation threshold is
$1/2$~\cite{NewmanStrogatzWatts2001, Newman2002}. Once $q$ exceeds $1/2$, a fault seeded in
the core has a positive probability, though not a certainty, of reaching a fixed fraction
of the core, and therefore of the whole graph. A small global $\kappa$ therefore does not
rule out propagation inside a dense component. Degree three alone does not produce this
result: a core made of separate groups of four changes that all contend with one another
also gives every core change three neighbours, yet a fault seeded there can never reach
more than four changes.

\emph{Adding a link can raise the threshold.} Let four changes all contend with one
another and sit beside two isolated changes, with degrees $(3,3,3,3,0,0)$. Then $\langle
k\rangle = 2$, $\langle k^2\rangle = 6$, $\kappa = 2$, and the reference threshold is
$1/2$. Adding one link between the two isolated changes gives degrees $(3,3,3,3,1,1)$,
$\langle k\rangle = 7/3$, $\langle k^2\rangle = 19/3$, $\kappa = 12/7$, and a reference
threshold of $7/12$. The added link joins two lightly contended changes, so a fault that
follows it meets no further links, and the average number of onward links falls. Removing
the link would lower the threshold again, although removing a link can never let a fault
reach more changes. The reference threshold is a ratio of moments and does not inherit the
monotonicity of the process it summarises.

\section{Following a Fault Link by Link: The Spectral Bound and Unequal Crossing Probabilities}\label{supp:operator}

\emph{In brief.} Equation~(1) uses only the degrees of a graph and assumes one crossing probability for every link. This section removes both simplifications. It follows a fault link by link, which leads to a matrix whose largest eigenvalue plays the role of $\kappa$. For equal crossing probabilities, this matrix is the non-backtracking matrix, which follows contention links in one direction without stepping straight back. The reciprocal of its largest eigenvalue is the \emph{spectral bound} of the main paper. The section also states what the bound does and does not guarantee. In particular, it is a statement about large sparse graphs and not a safety test for one finite graph. On a finite star, for example, the eigenvalue is zero, which suggests that no spread can sustain itself, yet a fault in the centre reaches every change when every link transmits. Finally, the section shows what changes when crossing probabilities differ across links or depend on contention degree.

Section IV-B refers to this section for the spectral bound and for unequal transmission
probabilities, which can be handled through a link-to-link matrix. Let $G$ be the contention graph on changes, and let each directed link
$j\to\ell$ carry a crossing probability $q_{j\ell}\in[0,1]$. The process starts from one
affected change, each directed link out of an affected change carries one independent
trial, and a change can be affected at most once.

Write $m_n(j\to\ell)$ for the expected number of times link $j\to\ell$ transmits in
generation $n$. A transmission along $j\to\ell$ in generation $n$ requires that $j$ became
affected in generation $n-1$ through some link $i\to j$ with $i\ne\ell$, and that the trial
on $j\to\ell$ succeeds. If the graph is locally tree-like, so that $\ell$ has not already
been reached by another route, the expectations satisfy, to first order,
\begin{equation}
m_n(j\to\ell) \;=\; q_{j\ell}\sum_{i:\,i\to j,\; i\ne \ell} m_{n-1}(i\to j),
\label{eq:supp:recursion}
\end{equation}
which is $\mathbf{m}_n = J\,\mathbf{m}_{n-1}$ for the column vector $\mathbf{m}_n$ indexed
by directed links, with $J_{(j\to\ell),(i\to j)} = q_{j\ell}\,\mathbf{1}\{\ell\ne i\}$ and
zero for pairs of links that do not chain. The row index is the link that transmits in
generation $n$, and the column index is the link that delivered the fault to its tail in
generation $n-1$. Counting generations from the first transmission, $\mathbf{m}_n =
J^n\mathbf{m}_0$, where $\mathbf{m}_0$, the vector of transmissions out of the origin $o$,
holds $q_{o\ell}$ on the links leaving $o$ and zero elsewhere. The exclusion $\ell\ne i$ removes immediate
backtracking, since under the at-most-once rule a change cannot pass a fault back to the
change it received it from. This linearisation is the first-moment argument behind
message-passing treatments of percolation~\cite{KarrerNewmanZdeborova2014}. It is exact on a
tree. Short cycles can allow repeated reach and make the tree approximation inaccurate.
Section IV-D compares the reference threshold with simulations but does not isolate this
source of error.

\emph{What the criterion states.} Three statements should be kept apart. (a) \emph{Finite
reach.} Every affected change other than the origin lies at the end of a non-backtracking
path from the origin on which every trial succeeded. On any graph, the walk counts therefore
bound the expected number of affected changes from above. The expected cascade size is at
most $1+\mathbf{1}^\top\sum_{n\ge 0}J^n\mathbf{m}_0$, a sum that converges when $\rho(J)<1$.
The bound can be large although $\rho(J)<1$: on a finite star, every non-backtracking walk
ends at a leaf, so $J$ is nilpotent and $\rho(J)=0$, yet at crossing probability one a
fault in the centre reaches every change. A small $\rho(J)$ is therefore not a guarantee of
small reach in a finite graph. (b) \emph{Instability of the linearised process.} When
$\rho(J)>1$ and the transmissions out of the origin reach the part of the graph that carries
the leading eigenvalue, the walk counts grow geometrically with the generation. This
describes the linearised branching process. It does not by itself establish large reach in
a given finite graph, since a change can be affected only once. (c) \emph{A giant cascade in a
growing graph family.} Consider a sequence of sparse, locally tree-like graphs whose size
grows. Suppose that the part of each graph that carries the leading eigenvalue holds a fixed
fraction of the changes and that the origin can reach that part. The message-passing
equations then become exact, and their trivial solution loses stability when $\rho(J)$
exceeds one. For homogeneous transmission, above that point a single fault reaches a fixed
fraction of the graph with positive probability. Where short cycles remain, or the leading eigenvalue is carried by a vanishing part of the graph,
$1/\rho(B)$ below is a lower bound on the percolation threshold rather than the threshold
itself~\cite{KarrerNewmanZdeborova2014, HamiltonPryadko2014}. The main
paper's reference threshold and spectral bound are statements of type (c).

\emph{Homogeneous transmission.} If $q_{j\ell}=q$ for every link, then $J = qB$, where $B$
is the non-backtracking matrix of $G$ with $B_{(j\to\ell),(i\to j)} = \mathbf{1}\{\ell\ne
i\}$. This $B$ is the transpose of the non-backtracking matrix as it is usually
written~\cite{KarrerNewmanZdeborova2014}. A matrix and its transpose have the same
eigenvalues, so the orientation leaves $\rho(B)$ and the bound below unchanged. The
criterion becomes $q>1/\rho(B)$, and $1/\rho(B)$ is the spectral bound of the main
paper, a lower bound on the percolation threshold of the given graph in the large sparse
limit~\cite{KarrerNewmanZdeborova2014, HamiltonPryadko2014}.

\emph{Homogeneous transmission on the configuration model.} With degree distribution $p_k$,
the change reached by following a random link has degree $k$ with probability proportional
to $k\,p_k$ and $k-1$ further links. Summing the recursion over the links reached, the
expected number of onward transmissions per transmission is
\begin{equation}
q\sum_k \frac{k\,p_k}{\langle k\rangle}(k-1) \;=\; q\,\frac{\langle
k^2\rangle-\langle k\rangle}{\langle k\rangle} \;=\; q\,\kappa,
\end{equation}
which is Equation~(1)~\cite{Newman2002, NewmanStrogatzWatts2001}. In this ensemble
$\rho(B)$ concentrates at $\kappa$ as the graph grows, so the spectral bound and the
reference threshold coincide in the limit and differ on finite graphs with degree
correlations or short cycles.

\emph{Unequal transmission.} If the $q_{j\ell}$ are draws with mean $\bar q$ that are
independent of one another and of the structure of $G$, the expected recursion is the
homogeneous one with $q=\bar q$, and only the mean enters. Correlated draws, such as a
common shock that raises every $q_{j\ell}$ at once, are not covered, because the expected
product along a path then need not equal the product of the means. When $q$ depends on structure, its effect can be
stated in the reference graph but not for every graph. Suppose that, on the
configuration model, the crossing probability of a link is a function $f(k)$ of the
contention degree $k$ of the change the link enters. A followed link reaches a change of
degree $k$ with probability $k\,p_k/\langle k\rangle$, crosses with probability $f(k)$, and
opens $k-1$ onward links, each of which reaches a change drawn again from this
link-weighted law. The mean-offspring matrix over degrees then has rank one, and the
branching ratio is
\begin{equation}
R_f \;=\; \sum_k \frac{k\,p_k}{\langle k\rangle}\,f(k)\,(k-1)
\;=\; \bar q\,\kappa + \operatorname{Cov}_{\mathrm{e}}\bigl(f(K),K\bigr),
\label{eq:supp:Rf}
\end{equation}
where $\bar q=\sum_k (k\,p_k/\langle k\rangle)\,f(k)$ is the crossing probability averaged over links, and $\operatorname{Cov}_{\mathrm{e}}$ is the covariance under the same
link-weighted law. If $f$ is nondecreasing and not constant over the degrees that occur,
the covariance is positive and $R_f>\bar q\,\kappa$, so an average $\bar q$ measured without
regard to degree understates the branching ratio. If $f$ is nonincreasing, $R_f\le\bar
q\,\kappa$. When $f$ rises with degree, averaging $q$ over changes rather than over links
gives a value no larger than $\bar q$, so the first conclusion also holds under that
convention, and the same expression~\eqref{eq:supp:Rf} results when $q$ depends on the
degree of the sending change instead. On a particular finite graph, the comparison with
$\bar q\,\kappa$ need not hold. On any finite tree, every non-backtracking walk ends at a leaf, so
$B$ is nilpotent and $\rho(J)=0$ whatever the $q_{j\ell}$, and on other graphs short cycles
and degree correlations separate $\rho(B)$ from $\kappa$. For a given large sparse graph,
the same linearisation makes $\rho(J)>1$ the natural extension of a criterion of type (c)
to unequal transmission. We have not derived the corresponding statement about the size of
the cascade, and the criterion is not an exact safety test for a finite, clustered, or
time-ordered graph.

A time-ordered version, in which a chain $(i\to j),(j\to\ell)$ is admitted only when the
exposure along $j\to\ell$ follows the exposure along $i\to j$, is not derived here.
Section~\ref{supp:temporal} gives exact calculations for one small schedule, which is
enough to show that order changes the prediction.

\section{Order on a Three-Change Schedule}\label{supp:temporal}

\emph{In brief.} The historical contention graph ignores the order in which changes merge. This section takes three changes that share one file and computes how many changes a fault is expected to affect. The same graph gives different answers depending on where the fault starts and on how exposure happens, which is why time order matters for the prediction.

Let change $A$ be open on days $[0,2]$, $B$ on $[1,4]$, and $C$ on $[3,5]$, all touching one
file and each merging at the end of its interval, so that the historical graph is the path
$A$--$B$--$C$. Let $q$ be the crossing probability per contention link. On the undirected
graph, ignoring time, a fault starting at either end affects $1+q+q^2$ changes in
expectation.

\emph{Exposure by branch update after a merge} (the channel of the episode in Section
III-C). A change can receive a fault only from a change that merges while it is open.
Starting in $A$, the fault can reach $B$, which is open when $A$ merges on day 2, and then
$C$, which is open when $B$ merges on day 4: the expected number of affected changes is
$1+q+q^2$. Starting in $C$, which merges on day 5 after $A$ and $B$ have closed, the fault
reaches neither, and the expected number is 1. Starting in $B$, it can reach $C$ but not
$A$, giving $1+q$.

\emph{Exposure at any time while both changes are open}, as through a shared workspace or a
test run against another open change. Starting in $A$, the fault can cross to $B$ during
$[1,2]$ and from $B$ to $C$ during $[3,4]$, giving $1+q+q^2$. Starting in $C$, it can cross
to $B$ during $[3,4]$, but $A$ closed on day 2, giving $1+q$.

At $q=1/2$, the expected cascade sizes from $A$ and from $C$ are 1.75 and 1 under the first
rule, and 1.75 and 1.5 under the second, on one graph with one degree sequence and one
reference threshold. The difference is produced by order alone, and it depends on which
exposure operations a repository uses. The structural measurements of the main paper are
therefore read as descriptions of opportunity, with their use in time order left to a
prospective construction that records exposures.

\section{Synthetic Model Checks in Detail}\label{supp:sim}

\emph{In brief.} We checked Equation~(1) on generated graphs, where the transition can be located by simulation. This section reports which graph families we generated, how well the method recovered thresholds that are known in closed form, and how the simulated transition, the spectral bound, and the reference threshold are ordered. The method recovered the known thresholds of Erd\H{o}s--R\'enyi graphs within 2\% and of random regular graphs within 0.5\%. However, it missed its 5\% criterion on a power-law family with exponent 3.5. The section closes by explaining why the generated graphs are much denser than real repositories, and why the gaps reported here are therefore not corrections for a repository's threshold.

This section reports the details that Section IV-D summarises. All values are from the
simulation summary and report in the replication package (68 combinations of ensemble and
graph size, 2,160 runs).

\emph{Ensembles.} Three calibration families have thresholds known in closed form:
Erd\H{o}s--R\'enyi graphs at three densities (15 combinations), power-law graphs (10), and
random regular graphs (10). Three ensembles are built from the corpus's distributions of
files per change and changes per file: the corpus-matched ensemble (5 sizes), a sweep over
the tail of the changes-per-file distribution (12), and variants with caps on how many
changes share one file (16). Sizes run from $N=1{,}000$ to $N=100{,}000$ changes.

\emph{Calibration.} Extrapolated to infinite size, the simulated transition recovered the
Erd\H{o}s--R\'enyi threshold with relative errors of $+0.66\%$, $+1.92\%$, and $+0.83\%$
at mean degrees 3, 6, and 10, against raw deviations of 2.36--2.93\% at $N=100{,}000$. It
recovered the random-regular threshold $1/(k-1)$ to within $+0.33\%$ and $+0.45\%$ at $k=4$ and $k=6$ (raw deviation 1.76\% at $N=100{,}000$). The estimator thus carries an upward bias of one to three percent
at the largest size even where the threshold is known. The power-law check had two legs.
At tail exponent $\alpha=2.5$, where the threshold falls toward zero as graphs grow, no
fixed value exists to compare with, and the simulation specification required instead that
the simulated transition decrease with $N$; it decreased strictly. At $\alpha=3.5$, the
heavy-tailed family whose second moment converges and whose threshold therefore stays above
zero as graphs grow, a fixed comparison is possible. There the extrapolated transition
exceeded the analytic tree-like value by $+6.41\%$ (interval $+2.78\%$ to $+10.02\%$,
$22.23\%$ at the largest size), against the specification's agreement criterion of 5\%.
This figure takes as its reference the analytic moments of the truncated power law that was
actually sampled. It extrapolates to an infinite graph with the finite-size exponent $1/5$,
the rate at which theory expects the transition to approach its limit for $3<\alpha<4$. Both
choices replaced those of a first pass and are recorded as deviations in the simulation
report. That leg failed, so part of any gap on heavy-tailed graphs may come from the
method.

\emph{Ordering.} In all 68 combinations the simulated transition lies above the reference
threshold. With the reference threshold computed from the pooled degrees of each
combination, as in Figure~2 of the main paper, the spectral bound lies between the two in
51 combinations, including all 33 built from the corpus's distributions. In the other 17,
seven Erd\H{o}s--R\'enyi and ten random regular combinations, it lies below the reference
threshold by less than 0.05\%, where theory places the two together. When the reference
threshold is instead averaged over runs, one further combination (tail sweep,
$\alpha=3.5$, $N=30{,}000$) has the spectral bound 8.5\% below it. The spectral bound
exceeds the simulated transition in 17 of the 2,160 runs, spread over $N=1{,}000$ to
$30{,}000$, and in no combination on average.

\emph{The corpus-matched ensemble.} The simulated transition exceeds the reference
threshold by $45.2\%$, $40.3\%$, $31.5\%$, $21.1\%$ (95\% CI 16.8--25.3), and $14.3\%$
(11.6--17.2) at $N=1{,}000$, $3{,}000$, $10{,}000$, $30{,}000$, and $100{,}000$. The spectral
bound exceeds the reference threshold by $24.2\%$, $21.7\%$, $15.0\%$, $9.6\%$, and
$4.79\%$ at the same sizes. Because the spectral bound is a lower bound on the percolation
threshold in the large-graph limit, this ordering places Equation~(1) below the threshold of
graphs of this kind independently of the simulation's error, to the extent that the limit
describes graphs of these sizes. The variants with caps on shared files agree
with the corpus-matched ensemble to within 2.6 points at the largest size. Extrapolating the
gap to an infinite graph returns $+9.3\%$, $+3.6\%$, or $-0.8\%$ depending on whether the
fit uses all five sizes or drops the smallest one or two. One cause is identifiable. At small $N$, a file of popularity $a$, that is, a file
shared by $a$ changes, cannot attach to $a$ distinct changes when $a$ is comparable to $N$.
The realised mean degree therefore reaches only 0.364, 0.565, 0.725, 0.851, and 0.946 of the 100.87 that the distributions imply. The sizes therefore mix the finite-size
shift of the transition with drift in the ensemble, and no single extrapolated value is
reliable.

\emph{Clustering as a possible cause.} Clustering offers one possible explanation for the gap
on the corpus-matched ensemble. All changes that touch one file form a fully connected group. A
cascade that enters the group can therefore reach some changes along several links, whereas
Equation~(1) counts each arrival as new. Repeated arrivals of this kind would raise the
simulated transition. We did not isolate clustering as the cause, because finite size, the
calibration error, and the drift of the ensemble with graph size also enter the comparison.

\emph{Distance from repositories.} The generated graphs preserve the corpus's distributions
of files per change and changes per file. They carry no temporal overlap, no repository
boundaries, and no correlation among the files a change touches, so every pair of changes
that shares a file is linked. Their mean degree runs from 36.7 at $N=1{,}000$ to 95.5 at
$N=100{,}000$ (85.88 at $N=30{,}000$), against 2.574 in the pooled corpus. The gaps above
are properties of these graphs. They are not corrections to apply to a repository's
reference threshold, and they do not show that the reference threshold is a conservative
safety bound for real repositories, whose dense cores and time order can move a transition
in either direction.

\section{Corpus Analyses in Detail}\label{supp:corpus}

\emph{In brief.} This section gives the details behind the corpus results in Sections III-B, V and VII of the main paper. It covers the comparison of messages with contending pairs, the file classes that mediate contention, and the fit of the tail of the degree distribution. It then compares the reference threshold with simpler measures, reports the test of modular structure and the share of each coding-agent product, and checks a collection limit of the corpus.

\emph{Provenance of the decision rules.} The decision rules of the message comparison, the
modularity test below, and the drift test of Supplement~S7 come from one written test
specification, dated 8 August 2026, the day on which the three analyses were run. The
specification was kept as a local document, not in an external registry, and the saved
file was last modified after the analyses, so the archive does not independently establish
that each rule preceded the corresponding result. The analysis report in the replication
package states each rule as it was applied.

\emph{Message comparison.} The primary ratio compares 4,674 messages posted by coding-agent
accounts on changes they did not author with the 42,600 contending pairs (0.110, 95\%
cluster-bootstrap interval 0.0645--0.233, 10,000 resamples by repository). The
specification's rule reads a ratio below 0.1 as support for the premise that agents in
this corpus meet mainly through shared code, a ratio of one or more as a refutation, and
values in between as a weakened premise. The two counts measure different things, so the
ratio compares how often each channel appears in the records. The specification lists
three sensitivity analyses. Counting generic automation accounts,
such as continuous-integration bots and coverage reporters, which are not coding agents,
raises the numerator to 26,225 and the ratio to 0.616 (0.331--1.389), an interval that
crosses one; this analysis changes the question, since those accounts do not author
changes. Restricting messages to those posted while the change was open gives 4,461 and
0.105 (0.0622--0.2225). Replacing the denominator by the number of contention incidences
counted with shared-file multiplicity, that is, each contending pair counted once for every
file the two changes share (448,457 in all, about 10.5 per pair), gives 0.0104
(0.0069--0.0167).

\emph{File classes.} We classified the files mediating contention as program source, other
source (markup, style, and query languages), configuration and manifest files,
documentation, and other assets. Of the contention links, 45.0\% are mediated by no
program-source file and 43.8\% by no source file of either kind. On program source alone,
ten of the thirteen repositories with reference thresholds below 5\% stay below 5\% and
hold 41.2\% of the changes; on source of either kind plus configuration, twelve stay and
hold 42.4\%. The pooled $\kappa$ falls from 107.6 to 25.7 and to 44.6, and the fitted tail
exponent rises from 2.12 to 3.06 (2.87--3.29) and to 2.81 (2.66--3.00). The program-source
restriction lowers
the reference threshold in 40 repositories and the source-plus-configuration restriction in
36. In the most contended repository, 96\% of contention runs through documentation and
continuous-integration files, and $\kappa$ falls from 283 to 25 on program source alone.

\emph{Tail fit.} A Clauset--Shalizi--Newman fit~\cite{Clauset2009} to the pooled degrees
selects $k_{\min}=38$ by the Kolmogorov--Smirnov criterion and gives $\alpha=2.12$ with 155
observations from 18 repositories in the tail. Resampling the tail degrees independently
gives 1.96--2.33. Resampling whole repositories, which respects the dependence among
degrees in one graph, gives 1.69--3.87 (5,000 resamples, $k_{\min}$ held fixed).

\emph{The reference threshold against simpler measures.} Table~\ref{tab:supp:proxies}
gives Spearman correlations between $\kappa$ and seven simpler measures. The largest
correlation among the 127 repositories with a threshold and at least thirty changes is 0.888 with the mean
contention degree. Across the repositories with a defined threshold, no measure places
more than six of the thirteen repositories below 5\% among its own top thirteen
(time-weighted open changes, concurrency per change, and mean contention degree each
recover six).

\begin{table}[!t]
\centering
\caption{Spearman correlation of $\kappa$ with simpler measures, with 10,000-draw
bootstrap intervals.}
\label{tab:supp:proxies}
\scriptsize
\setlength{\tabcolsep}{3pt}
\setlength{\tabcolsep}{2pt}
\begin{tabular}{@{}lll@{}}
\toprule
\textbf{Measure} & \textbf{472 repositories} & \textbf{127 repositories} \\
\midrule
Agent-authored changes & 0.554 [0.481, 0.619] & 0.285 [0.099, 0.449] \\
Open changes, time-weighted & 0.518 [0.445, 0.584] & 0.647 [0.522, 0.748] \\
Concurrently open per change & 0.712 [0.656, 0.762] & 0.691 [0.567, 0.786] \\
Mean contention degree & 0.752 [0.703, 0.794] & 0.888 [0.825, 0.928] \\
Distinct files touched & 0.452 [0.375, 0.522] & 0.336 [0.169, 0.481] \\
Distinct agent products & 0.099 [0.010, 0.187] & $-0.004$ [$-0.179$, 0.174] \\
Distinct accounts & 0.109 [0.015, 0.200] & 0.079 [$-0.090$, 0.243] \\
\bottomrule
\end{tabular}
\end{table}

\emph{Modular structure.} Section VII of the main paper reports that the specified
test of whether modular structure limits contention did not support it.
Table~\ref{tab:supp:arch} gives the detail. The primary analysis uses the 134 repositories
with at least thirty changes, 127 of which carry a threshold, and repeats at fifty and a
hundred changes. Three separate regressions each relate the standardised $\log(1+\kappa)$
to one standardised index of the directory tree the changes touched: breadth across
top-level directories, depth, or the share of files at the root. Each controls for the
logged numbers of changes and of distinct files, with HC3 standard errors and Holm
correction across the three indices. The specification's rule required at least two
indices with an effect in the predicted direction at Holm-corrected $p<0.05$ and none
significant against it. At the primary cut it was not met: depth carried the predicted
negative coefficient, breadth a positive coefficient against the predicted negative sign,
and the share of files at the root, predicted positive, showed no detectable association.
The indices describe the subtree that the corpus's changes touched, not the repository's
architecture, and broad top-level spread is also what a monorepo looks like.

\begin{table}[!t]
\centering
\caption{Standardised coefficients of the modularity regressions (Holm-corrected $p$ in
parentheses). Predicted signs: breadth negative, depth negative, root share positive.}
\label{tab:supp:arch}
\footnotesize
\setlength{\tabcolsep}{3pt}
\begin{tabular}{@{}lrrr@{}}
\toprule
\textbf{Cut} & \textbf{Breadth} & \textbf{Depth} & \textbf{Root share} \\
\midrule
$\ge$30 changes (134) & $+0.250$ (0.015) & $-0.214$ (0.022) & $+0.020$ (0.89) \\
$\ge$50 changes (83) & $+0.287$ (0.004) & $-0.270$ (0.024) & $+0.178$ (0.049) \\
$\ge$100 changes (32) & $+0.207$ (1.00) & $-0.102$ (1.00) & $-0.023$ (1.00) \\
\bottomrule
\end{tabular}
\end{table}

\emph{Product shares.} One coding-agent product authored 65.8\% of the corpus's changes,
95.9\% of the changes in the thirteen repositories below 5\%, and 80.0\% of the changes in
the fitted tail.

\emph{Commit ceiling.} The pinned revision of AIDev records at most thirty commits per pull
request, because its collection did not page through longer lists. The dataset's maintainers
document this limit and removed it in a later release. File lists are built from the recorded
commits, so a change with more than thirty commits may lack some of its files and, with them,
some contention links. In the corpus, 227 changes carry exactly thirty recorded commits and
none carries more. As a check, we removed these 227 changes and their links and recomputed
every threshold. The corpus then holds 32,878 changes and 41,145 contention links, and 465
repositories have a defined threshold, with a median of 56\%. Twelve of the thirteen
repositories stay below 5\% and contain 42.4\% of the changes, and the pooled $\kappa$ is
110.5. The repository that leaves the group contained one of the removed changes. The check
removes changes that may be incomplete, and it does not recover missing files. The same limit
applies to the comment table, in which 44 changes carry exactly thirty comments, so the
message count above is a lower bound for this reason as well.

\section{Transmission-Estimation Pilot: Protocol and Attrition}\label{supp:pilot}

\emph{In brief.} We attempted to measure the crossing probability directly by seeding faults in twenty pairs of contending changes. This section documents the protocol and its versions, and it explains why none of the twenty pairs produced a usable measurement. It also states what this attrition does and does not imply.

Section VIII summarises a pilot that attempted to measure $q$ by seeding faults in
contending pairs drawn from the version histories of public repositories. Seeding faults is one route to $q$, not the
only one. Identification from observational data is possible in principle under stated
causal assumptions, but the corpus lacks the provenance of faults, the outcomes of
exposures, and the temporal order those assumptions would need. This section records the
protocol, its versions, and the attrition of its final run (Stage~1b); the protocol, its
amendments, the stage reports, and the per-pair records are in the replication package.

\emph{Versions, and what the archive establishes.} The protocol was kept as a locally
versioned document. It was not deposited in an external registry, and no independent
timestamp of its earlier versions exists. Version 1.0 was approved at a first gate
(Stage~0) on 29 July 2026. The amendment record dates four amendments and two corrections
to that gate. A fifth amendment, which split the injection procedure by change type, is dated
to the reconstruction of the twenty pairs, before any fault was seeded. A first run
(Stage~1), the initial instrument check of Section VIII, then reached nine terminal states, which met the stopping rule's minimum
of eight, but produced no verified escape measurement. Four further amendments repaired
the instrument after that attrition was known: syntax-aware discovery of mutation sites,
build images pinned to each repository's own, subtypes for merge conflicts, and a frame
weight for repositories without tests. The final run, Stage~1b, used the same twenty pairs
without resampling. The protocol and amendment files were last saved a few minutes after
the Stage~1b run ended, and they contain notes on its outcome. They therefore record the
protocol as it stood after Stage~1b and do not by themselves establish which wording
preceded which trial. The Stage~1 report, saved before Stage~1b began, already treats the
five operator classes below as fixed by the protocol. No file in the archive that was
saved before Stage~1b records the second and third criteria of the gate described below.
The package holds these documents with their wording edited for readability, together with
a provenance note that records their dates; methods, numbers and decisions are unchanged.

\emph{Protocol.} Twenty contending pairs $(c, c')$ of agent-authored changes, ten Python and
ten TypeScript, were sampled from a worklist drawn from the corpus under a cap of twenty
trials per repository, with $c$ and $c'$ overlapping in time and both modifying a common
non-test source file. Each pair was reconstructed in a container at the merge base
of the two heads. A fault was seeded in the region $c$ modified, using one of the
protocol's five operator classes (arithmetic operator, conditional boundary, negation
removal, argument swap, returned value). The injection procedure differed by change type: for
a change that modifies existing lines, the
fault was injected into the merge base at that region; for a change that only adds lines,
$c$ was applied first and the fault injected within its added lines. Change $c'$ was then
applied, and the repository's own verification was run with $c'$ applied, before and after
the mutation. A trial yields an escape measurement when the baseline passes, the tests are
runnable, the mutation survives the application of $c'$, and the outcome after mutation is
recorded. Dependence was judged by coverage of the mutated line where a suite ran, with
static reachability recorded alongside. Every trial ends in a terminal state (a crossing,
no dependence, a dependence that verification detected, a mutant that no test reaches, or a
merge conflict) or in a named exclusion. The Stage~1b report applies a gate that required,
in twenty pairs, at least eight terminal states, at least four measured escapes on passing
baselines, and at least two coverage-based dependence judgements before a larger stage
could begin. The project records describe this gate as set before the re-run; the archive
cannot confirm that independently.

\emph{Attrition.} Table~\ref{tab:supp:funnel} gives the funnel. Of twenty pairs, seventeen
were reconstructed: one repository was no longer public, and in two the first change no
longer touched the candidate file at the merge base. Of the seventeen, three failed to
compose $c'$ on $c$ before any fault was seeded, seven offered no site for any of the five
operator classes once comments, strings, type annotations, and markup were excluded, and
seven yielded a site. Of those seven, three collided with $c'$ at the mutated lines, two
had a verification suite that failed before mutation, and two had no runnable tests. The
two trials without runnable tests were recorded as bounded crossings because the protocol
assigns escape by construction where nothing verifies, and they are not measured
transmission events. The gate's first criterion was met at exactly its threshold, by six
merge conflicts and the two bounded crossings, and its
second and third returned zero, so the study stopped under that gate.

\begin{table}[!t]
\centering
\caption{Attrition of the pilot, final run (Stage~1b, twenty pairs).}
\label{tab:supp:funnel}
\footnotesize
\begin{tabular}{@{}lr@{}}
\toprule
\textbf{Stage and outcome} & \textbf{Pairs} \\
\midrule
Sampled & 20 \\
\quad lost at reconstruction (repository unavailable) & 1 \\
\quad lost at reconstruction (file not touched at merge base) & 2 \\
Reconstructed & 17 \\
\quad composition of $c'$ on $c$ fails & 3 \\
\quad no site for any of the five operator classes & 7 \\
\quad site found and fault seeded & 7 \\
\qquad mutation collides with $c'$ & 3 \\
\qquad verification fails before mutation & 2 \\
\qquad no runnable tests (escape assigned by construction) & 2 \\
\qquad passing baseline, runnable tests, surviving mutation & 0 \\
Measured escapes & 0 \\
Coverage-based dependence judgements & 0 \\
\bottomrule
\end{tabular}
\end{table}

\emph{What the attrition implies.} Zero measured trials in twenty attempts give $0.168$ as
the upper endpoint of the two-sided 95\% Clopper--Pearson interval for the per-pair yield
of this instrument; the one-sided 95\% upper bound is $0.139$. The pairs were selected,
stratified by language, and capped per repository, and the instrument was revised between
the two runs on the same pairs, so the interval describes this instrument on this worklist
rather than a rate in a wider population. The following figures are therefore illustrative
yield planning under stated assumptions, not a validated resource forecast, and they say
nothing about the value of $q$. At a yield of $0.168$, four hundred measured trials would
require about 2,375 eligible pairs, against 2,578 in the capped worklist. At a yield of 5\%,
they would require 8,000, which the uncapped frame of 8,972 eligible pairs in the two
languages could supply only by breaching the per-repository cap. At 2\%, they would require
20,000, which the frame cannot supply. Neither time (284 seconds across seventeen trials) nor disk (at most
3.9~GB for one trial) came near its budget. The binding constraint was the conjunction the
design requires: a reconstructible pair, a mutable site under the fault model, a clean
composition, a passing baseline, and a runnable suite. In these seventeen pairs, the
regions the first change had modified were mostly imports, type declarations,
configuration literals, and one-line edits. Widening the fault model to statement
deletion, literal replacement, or identifier substitution could reach such regions, and
accepting a type check or lint gate as verification where a suite is absent could convert
some failing or test-less baselines. Either change alters what is measured and requires a
revised protocol, and neither is certain to raise the yield.

The design has a further limit that a higher yield would not remove. It applies $c'$ as it
was recorded, so even a successful trial measures whether a seeded fault survives
composition with a fixed change and is missed by verification. In the episode of Section
III-C, the receiving change incorporated the fault because its agent responded to a failing
test. Measuring that response requires agents that act after exposure, as in a controlled
study with live agents. The gate failed, so the pilot shows that this instrument, applied
to this frame, did not yield a measurement; it says nothing about the value of $q$ or about
whether faults cross contention links.

\section{Predictions T2 and T3 in Detail}\label{supp:predictions}

\emph{In brief.} This section supports predictions T2 and T3 of the main paper. For T2, it reports diagnostics of how $\kappa$ changes with repository size. For T3, it outlines what must be derived before the scaling prediction can be tested.

\emph{The T2 size diagnostic.} Each repository's degrees were sampled independently from one
corpus-level law (empirical below $k_{\min}=38$, power law with $\alpha=2.118$ above),
truncated at $n-1$ for a repository of $n$ changes, in 1,000 replicates. Sampled degrees
need not form a contention graph, so the comparison is a diagnostic of degree
distributions, not of graphs. The null band already contains the growth of $\kappa$ with
size that a fixed heavy-tailed law produces. Across six repository-size groups (boundaries
at 6, 21, 55, 144, and 3,569 changes), the observed $\kappa$ exceeds the null's 97.5th
percentile only in the group of 55--144 changes (19.99 against 18.87) and lies inside or
below the band in the other five.

\emph{The drift test.} The test in the specification described in Supplement~S5, whose
provenance note applies here, compared tail exponents across three size strata, each
required to hold at least 100 tail observations. The corpus carries 155 tail
observations in all, so no division into three strata can give each of them 100. The
smallest stratum, the 11,046 changes in repositories of one to 42 changes, contains no
change with a contention degree of 38 or more; only 406 of its changes sit in repositories
large enough to reach that degree. A power analysis with the same machinery, holding the
corpus's mix of repository sizes fixed, found zero power at the current corpus size and at
three, ten, and thirty times that size, for drifts of 0.2 and 0.4 in the exponent. In its
simulations the smallest stratum never reached 100 tail observations, averaging 0.15 at the
current size and 4.1 at thirty times it. In the corpus, 59.4\% of repositories (1,621 of 2,731) hold two
or fewer agent-authored changes.

\emph{A narrowing rate.} A closed form for how fast the interval below the reference
threshold narrows with size can be written from the fitted tail. We do not state one,
because its exponent is 0.79 on the primary graph, 0.10 on source plus configuration, and
negative on program source alone, and a rate whose sign changes with defensible
choices of how the graph is built is not a robust result.

\emph{T3 and the scaling programme.} At the transition, mean cascade size grows without
bound in a large graph. For finite repositories, T3 predicts that cascade-size
distributions rescaled by powers of repository size fall onto one curve. Consider a
critical branching process seeded with one affected change. Its offspring law is the
distribution of the number of new affected changes that one affected change produces. If
that law has finite variance, the probability that the cascade affects $n$ changes in all,
its total progeny, decays as $n^{-3/2}$, a line of slope $-3/2$ on logarithmic axes~\cite{Otter1949}. The statement
concerns this seeded process; it describes neither the counts of all clusters in a graph
nor every finite contention graph. In the reference graph, the origin and the later generations follow different offspring laws. A uniformly seeded origin
has degree $k$ with probability $p_k$ and produces a Binomial$(k,q)$ number of offspring. A
change reached along a link has degree $k$ with the size-biased probability $k\,p_k/\langle
k\rangle$ and produces a Binomial$(k-1,q)$ number, so the later generations have mean
$q\kappa$ and variance
\begin{equation}
q(1-q)\,\kappa + q^2\Bigl(\frac{\langle k^3\rangle}{\langle k\rangle} -
\frac{\langle k^2\rangle^2}{\langle k\rangle^2}\Bigr).
\end{equation}
For a fixed $q>0$, this variance is finite only if the third moment $\langle k^3\rangle$ of
the contention degrees is finite; a finite variance of the contention degrees does not
suffice. The reference threshold is not a fixed $q$, however. It falls as larger
repositories reveal more heavily contended changes, and whether the variance at
$q=q_c$ stays bounded then depends on how $q_c$ and the degree moments scale together with
the largest degree, which we have not derived for the corpus's distribution. Whether a
repository's offspring law has finite variance in the relevant limit is therefore not
settled. The rescaling exponents must be derived for the specified graph family before a
collapse can be tested, since a collapse achieved under freely adjusted exponents would
test nothing.

\section{Causal Emergence: Worked Example and Decision Rule}\label{supp:twosite}

\emph{In brief.} Section VIII proposes to compare a detailed and a grouped description of the same faults by their effective information. This section works the smallest example. Two code locations can each be correct or faulty. This gives four detailed states (neither location faulty, the first, the second, or both) and three grouped states (zero, one, or two faults). Intervening evenly over the four detailed states makes a single fault twice as common as no fault or two faults, whereas intervening evenly over the three counts gives each count one third. When the outcome depends only on how many locations are faulty, the grouped description scores higher under its own weighting. Under one common weighting, however, it cannot carry more than the detailed description. The section ends with the decision rule for proposition E1.

This section works the example of Section VIII under one explicit response rule. Two code
locations are each correct ($0$) or faulty ($1$). The detailed description has four states: $00$, $10$, $01$, $11$. The grouped description counts faults and has three states: $0$, $1$, $2$. Effective information, following Hoel et al.~\cite{Hoel2013, Hoel2017}, is the
mutual information between a cause state drawn uniformly from the states a description
distinguishes and the effect state one cycle later.

Take the rule that a fault at either site spreads to both within one cycle, and that a
fault-free pair stays fault-free. The detailed transitions are $00\to00$, $10\to11$,
$01\to11$, $11\to11$, and the grouped transitions are $0\to0$, $1\to2$, $2\to2$. All are
deterministic, so the conditional entropy of the effect given the cause is zero, and
effective information equals the entropy of the effect distribution.

\emph{Detailed description under its own uniform intervention.} The four causes are imposed
with probability $1/4$ each. The effect is $00$ with probability $1/4$ and $11$ with
probability $3/4$, so the effective information is $H(1/4,\,3/4) = 0.811$ bits.

\emph{Grouped description under its own uniform intervention.} The three counts are imposed
with probability $1/3$ each, and imposing a count of one means choosing $10$ or $01$ at
random. The effect is $0$ with probability $1/3$ and $2$ with probability $2/3$, so the
effective information is $H(1/3,\,2/3) = 0.918$ bits.

\emph{Detailed description under the grouped weights.} Imposing the detailed states with
the weights the grouped intervention induces, $00$ and $11$ with probability $1/3$ each and
$10$, $01$ with probability $1/6$ each, gives an effect distribution of $00$ with
probability $1/3$ and $11$ with probability $2/3$, and an effective information of $0.918$
bits.

The grouped description gains $0.107$ bits under the comparison that intervenes evenly over
each description's own states, because the two detailed causes $10$ and $01$ produce the
same effect and the detailed intervention spends half its weight on them. Under the grouped
weights the detailed description carries exactly as much information as the grouped one,
as it must, since it distinguishes everything the grouped description distinguishes. The
gain therefore reflects how the two descriptions weight the underlying configurations under
intervention, and implies no information absent from the detailed state. Under a rule in
which the two single-fault states produced different effects, the detailed description
would carry more under its own weights, and the grouped description would show no gain.

\emph{Decision rule for E1.} A positive result requires a simultaneous 95\% interval on the
gain in effective information that lies above zero. Before any repository is measured, the
estimators must recover known values on constructed transition rules. A rule that keeps
each group's fault count while placing faults at random within the group must show a
gain. A rule that copies the detailed state unchanged, or sends every state to the
same successor, must not. A precise interval that excludes a gain rejects E1 for the
groupings, contexts, and intervention mixtures tested. A gain in only some groupings makes
E1 conditional on the grouping, and an interval too wide to decide leaves it open, so a
null estimate without adequate precision does not refute it. Every grouping declared in
advance is reported. Information decomposition, which splits what several variables jointly
say about an outcome into unique, redundant, and synergistic parts~\cite{Rosas2020}, answers
a different question on observational data and serves as a secondary check. Agreement strengthens a result, and disagreement implicates the intervention distribution, the
grouping, or the estimators without deciding among them.

\section{Validation Studies}\label{supp:next}

\emph{In brief.} This section outlines four validation studies that follow from the paper. These are a demonstration that an agent incorporates a seeded fault (V1), a test of Equation~(1) on observed graphs (V2), a comparison with human-authored changes (V3), and the outcomes that a transition test should record (V4). It then states how a test of T1 is read, the decision rule for C1, and why agent adaptation bears on the crossing probability.

The paper's evidence is structural, and the first studies it needs are designed here in
outline. None has been run, and each would be pre-registered before data collection.

\emph{V1, incorporation demonstration.} The question is whether a coding agent that updates
its branch onto a merged change carrying a seeded fault builds that fault into its own
change. Each constructed scenario is a small executable repository with a specification,
visible tests, hidden oracle tests derived from the specification, and a continuous
integration script. A merged source change carries one seeded fault that its own tests do
not catch. The agent receives a task that modifies the same file and depends on the faulty behaviour. The harness updates the agent's branch onto the faulty main branch at a
fixed point, so that exposure is logged and incorporation is measured conditional on it.
Incorporation is recorded when an oracle test of the agent's own functionality fails and
still fails after the source's fault site is restored to its correct version. Six scenario
families vary the fault (a boundary rule as in Section III-C, units, rounding, a
configuration default, a documented contract, an error contract), each in Python and
TypeScript. Controls are a correct source change, a task that shares the file without
depending on the fault, and a task that depends on the fault through a call from another
file, the route outside the contention graph. A further arm without a visibly failing test
separates incorporation driven by test repair from silent dependence, and a second
generation asks whether a later task inherits the error from the receiving change. Three
pinned agent configurations and ten runs per scenario give 120 trials per arm and agent
(60 per language). In a one-sided Fisher test at the 5\% level, 120 trials detect a rise
from 5\% to 20\% incorporation with power of about 0.97, and 60 trials detect a rise from
5\% to 25\% with power above 0.9. An instrument stage
with a pre-registered continuation gate precedes the full design. A positive result
demonstrates the mechanism in constructed scenarios and estimates the incorporation factor
of $q$ there; it does not estimate the population's $q$ or test T1.

\emph{V2, Equation (1) on observed graphs.} Three quantities are computed for each of the 127
repositories with at least thirty changes and a defined threshold: the reference threshold,
the spectral bound, and the simulated transition. Each is computed on the observed contention
graph, on 100 degree-preserving rewirings of it, and on a time-respecting version in which
links run from a change to the changes still open when it merges. Comparing rewired with reference graphs isolates the tree approximation and finite size, comparing observed with rewired graphs isolates structure beyond the degrees, and comparing time-respecting with observed graphs isolates order. Because many graphs are
small, the probability that a cascade reaches at least a tenth of the changes is reported
as a function of $q$ beside the susceptibility peak.

\emph{V3, human-authored comparison.} Human-authored pull requests over the same
repositories and window give human-only, agent-only, and mixed contention graphs. The
comparison is descriptive. If agent-authored work produces higher $\kappa$ or more
concentrated contention at matched volume, the concentration is associated with agent
authorship in these repositories. Matching volume does not remove differences in task mix,
product, repository, or exposure, however, so the comparison would not by itself show that
agent authorship causes the concentration. If agent-authored work differs by less than a declared margin at
matched volume, the concentration is better described as a property of busy repositories.

\emph{V4, outcomes for a transition test.} A fault reaching one further change is a
transmission and occurs below the threshold too. A transition is read from the extent of
spread: mean and largest cascade size per origin, cascade depth, susceptibility, and the
probability of reaching a declared fraction of open changes, each recorded as $q$ varies.

\emph{Reading a test of T1.} The transition is read from a statistic declared in advance, such
as the susceptibility of Section IV-D as $q$ varies. Departures in clustering, degree
correlation, and time order are measured beforehand, by comparing the graph with its
degree-preserving rewirings and with its time-respecting version (V2). On a large sparse graph,
a transition below the spectral bound would point to a failed transmission assumption or a
measurement error, because the spectral bound is a lower bound on the percolation threshold in
that setting (Section~\ref{supp:operator}).

\emph{Decision rule for C1.} A propagated-fault event is a fault that reaches the main branch
in a change other than the one in which it began. The term names a transmission outcome only,
and it claims no broader loss of control. The outcome of the comparison in Section VIII is the
number of such events under the matched budget, with a minimum benefit declared in advance. A
structural advantage is shown only by an interval that lies above that margin. Equal
performance is shown only by an interval within a declared equivalence margin, and never by a
nonsignificant difference.

\emph{Agent adaptation and the crossing probability.} Section III-D of the main paper treats the single, fixed $q$ as a simplifying assumption, because adaptation can change each part of the
model. A changed response can move any factor of $q$. In the episode of Section III-C, the
agent working on $e_2$ moved incorporation when it repaired a failing test by matching the
code. A new or repurposed route changes where exposures can occur. A value of $q$ fitted in
one context may therefore not hold after agents adapt. For this reason, the comparison of
transmission before and after agents repurpose a connection is among the larger tests listed
below.

Larger tests lie beyond the paper: calibration across the four kinds of connection,
prediction in configurations not used for fitting, the behaviour of calibrated
transmission when agents adapt, prospective forecasting, and transfer to industrial
settings.

\bibliographystyle{IEEEtran}
\bibliography{bib}